\documentclass[12pt,preprint]{aastex}
\def\Sec{${}^{\prime\prime}$\llap{.}}

\def\teff{T$_{\rm eff}$}
\def\logg{log $g$}
\begin{document}
\title{Carbon and Strontium Abundances of Metal-Poor Stars\altaffilmark{1}}

\author{David K. Lai}
\affil{University of California Observatories/Lick Observatory, Department of Astronomy and Astrophysics, University of California, Santa Cruz, CA.}
\author{Jennifer A. Johnson}
\affil{Department of Astronomy, Ohio State University, Columbus, OH.}
\author{Michael Bolte}
\affil{University of California Observatories/Lick Observatory, Department of Astronomy and Astrophysics, University of California, Santa Cruz, CA.}
\and
\author{Sara Lucatello}
\affil{Osservatorio Astronomico di Padova, Vicolo dell'Osservatorio 5, 35122 Padua, Italy.}

\email{david@ucolick.org, jaj@astronomy.ohio-state.edu, bolte@ucolick.org, sara.lucatello@oapd.inaf.it}

\altaffiltext{1}{The data presented herein were obtained at the W.M. Keck Observatory,
which is operated as a scientific partnership among the California
Institute of Technology, the University of California and the
National Aeronautics and Space Administration. The Observatory was
made possible by the generous financial support of the W.M. Keck
Foundation.}


\begin{abstract}

We present carbon and strontium abundances for 100 metal-poor stars
measured from R$\sim $7000 spectra obtained with the Echellette
Spectrograph and Imager at the Keck Observatory.  Using spectral
synthesis of the G-band region, we have derived carbon abundances for
stars ranging from [Fe/H]$=-1.3$ to [Fe/H]$=-3.8$.  The formal errors
are $\sim 0.2$ dex in [C/Fe].  The strontium abundance in these stars
was measured using spectral synthesis of the resonance line at 4215
{\AA}.  Using these two abundance measurments along with the
barium abundances from our previous study of these stars, we show it
is possible to identify neutron-capture-rich stars with our spectra. We
find, as in other studies, a large scatter in [C/Fe] below [Fe/H]$ =
-2$.  Of the stars with [Fe/H]$<-2$, 9$\pm$4\% can be classified as
carbon-rich metal-poor stars. The Sr and Ba abundances show that three
of the carbon-rich stars are neutron-capture-rich, while two have
normal Ba and Sr. This fraction of carbon enhanced stars is consistent
with other studies that include this metallicity range.

\end{abstract}


\section{Introduction}\label{intro}

The study of metal-poor stars has evolved in the past twenty years
from the close examination of the few known stars with [Fe/H] $<-3.0$
to the investigation of the statistical properties of large samples of very
metal-poor stars. This new wealth of stars with low
[Fe/H], originally discovered in the surveys of \citet{bond80} and \citet{bps}
(hereafter, BPS), and more recently in new surveys, including the
Hamburg/ESO Survey \citep{hes} and SDSS \citep{sdssmetpoor}, has revealed a
number of subgroups among the metal-poor stars characterized
by different abundance patterns. These include
stars with large enhancements of elements made by the r-process
(e.g.,\citealt{mcwilliam95,hill02}) and stars with large enhancements in
the $\alpha$-elements \citep{aokialpha}. Among the most prominent 
types are the carbon-rich objects (here we define C-rich as [C/Fe]$\geq$1).
As increasing numbers of C-rich stars were studied, it became clear that
the abundances of other elements could vary widely from one C-rich star
to another. \citet{beers05} classified these objects into four subclasses
based on the abundances of the heavy neutron-capture elements Ba and Eu: 
CEMP-r, CEMP-s, CEMP-r/s and CEMP-no. The CEMP-r objects have enhanced
Ba and a low [Ba/Eu] ratios that suggest solely a r-process contribution,
while the CEMP-s and CEMP-r/s also have enhanced Ba, but higher [Ba/Eu]
ratios which suggest that the s-process has polluted them only (-s)
or as well (-r/s).
The CEMP-no stars show no enhancement in Ba or Eu.

One of the goals of current research on metal-poor stars is to
quantify the ratios of the different types of C-rich stars as a function of
[Fe/H]. This will isolate those phenomena associated with the
early Universe and test whether certain types of events were more
common. An example of the former are the extremely large [C/Fe]
abundances found in the two most [Fe/H]-poor stars
\citep{christlieb02,frebel05}. An example of the latter may be the
higher fraction of stars at low [Fe/H] that show signs of carbon mass
transfer from an asymptotic giant branch (AGB) companion.  

Another
important quantity is the the overall fraction of C-rich stars, as a
measure of the frequency of carbon pollution events. The percentage of
C-rich stars increases as [Fe/H] decreases
\citep{lucatello06,rossi05}, going up to 100\% at
[Fe/H]$<-5$. However, the total fraction of CEMP stars at low
metallicities is in dispute. Estimates range from 25\% for
[Fe/H]$<-2.5$ \citep{marsteller05} to $14\pm4$\% for [Fe/H]$<-2.0$
\citep{cohen05} to $9\pm2$\% for the same metallicity range in
\citet{frebel06}. \citet{lucatello06} found a lower limit of
21$\pm2$\% for the fraction of stars with [Fe/H] $<-$2.0 that are
C-rich in the HERES sample \citep{heres}. Possible reasons for the
discrepancies between studies include underestimates in some surveys
of the [Fe/H] for stars with strong CH/CN molecular features
\citep{cohen05}, trends in the fraction of C-rich stars with distance
from the Galactic plane \citep{frebel06}, and the inclusion in samples
of giants whose C abundance has been decreased because evolutionary
dilution \citep{lucatello06} and/or has been processed some of their
carbon through the CN cycle \citep{aoki07}.  Therefore, increasing the
sample size of metal-poor stars with C-abundances, especially for
stars with warmer atmospheres that are the least likely to suffer from
most of the above effects, is important for improving statistics at
the edge of the metallicity distribution function.

The other goal of this study is to further constrain the evolution of
neutron-capture elements, specifically by comparing measured
abundances for the light neutron-capture element Sr, with the heavy
neutron-capture element Ba.  By examining the light neutron-capture
elements, Sr-Y-Zr, it has been shown that there must be a
distinct neutron-capture process besides the classical s- and
r-processes that contributes to their abundances (e.g. \citealt{mcwilliam98})
 A likely source for
some of this production is the weak s-process in more metal-rich stars
(e.g. \citealt{truran,burris2000}), but 
to fully understand the abundances, \citet{travaglio} suggest that
there needs to be another production site which is unrelated to the
weak s-process that they call LEPP (lighter element primary process).  
It is important to
obtain more measurements of [Sr/Ba] to further constrain the magnitude
of these processes, as well as where it provides the greatest contribution.
This is particularly relevant for the CEMP stars, because, as mentioned
previously, they display a wide variety in the abundance ratios of the
heavy elements.

Previous high-resolution studies, e.g. \citet{mcwilliam98},
\citet{burris2000}, \citet{aoki05}, and \citet{heres}
explored this ratio for a number of stars in the metal-poor regime.
 In this study we show it
is possible to obtain this ratio from relatively low-resolution
spectra.

We report on the carbon and strontium abundances for 100 stars,
which are most
of the sample of \citet{lai04} (hereafter, Paper I). While this sample is
 biased toward
metal-poor stars, it does not suffer from substantial biases in [C/Fe]
ratios among the warmer stars.  It extends to higher metallicities
than other studies, which allows us to observe trends in number of
C-rich stars with [Fe/H]. We can pair [C/Fe] and [Sr/Fe] measurements
with the Ba
abundances from Paper I and place these stars into their CEMP
subclasses. Preliminary measurements of the carbon abundances for this
sample were used to identify two stars for higher resolution
follow-up: CS~22183-015 \citep{jbcs22183} and CS~31062-050
\citep{jbcs31062}.

\section{Observations, Sample Selection and Reductions}

The details of the observations and reduction procedures are
presented in Paper I; we present only an outline here.  
The goal of Paper I was to measure [Fe/H] and abundance ratios
in a large number of metal-poor candidates. The candidates were
selected from four sources: \citet{norris99},
\citet{twarog00}, \citep{int} and
the original BPS paper.  We attempted to select the most metal-poor
candidates from each survey. For \citet{twarog00}, \citet{int}, and BPS,
we selected candidates based on their [Fe/H] estimates. For
\citet{norris99}, our selection was based on the UBV excess.
For this paper we restrict our carbon measurements to the
objects from Paper I with [Fe/H] $< -1.3$ 
because the metallicity and atmosphere
determinations are uncertain for higher metallicities. We note that
the spectrum for CS30332-067 was misidentified in our Paper I,
and is therefore excluded here. This yielded a
total of 100 objects for this study. 

We obtained our spectra over the course of several runs using the
Echellette Spectrograph and Imager (ESI)
\citep{sheinis02} at the Keck 2 telescope
We ran
ESI in the echellette mode with the 0\Sec75-wide slit, resulting in a
spectral resolution of $R \simeq 7000$.  The wavelength coverage of
the spectra easily encompasses the G-band region used in our
analysis. We then used IRAF\footnote{IRAF is distributed by the
National Optical Astronomy Observatories, which are operated by the
Association of Universities for Research in Astronomy, Inc., under
cooperative agreement with the National Science Foundation.} to
extract the spectra to one-dimensional, wavelength calibrated,
continuum-divided form.

\section{Abundance Analysis}

\begin{figure}
\plotone{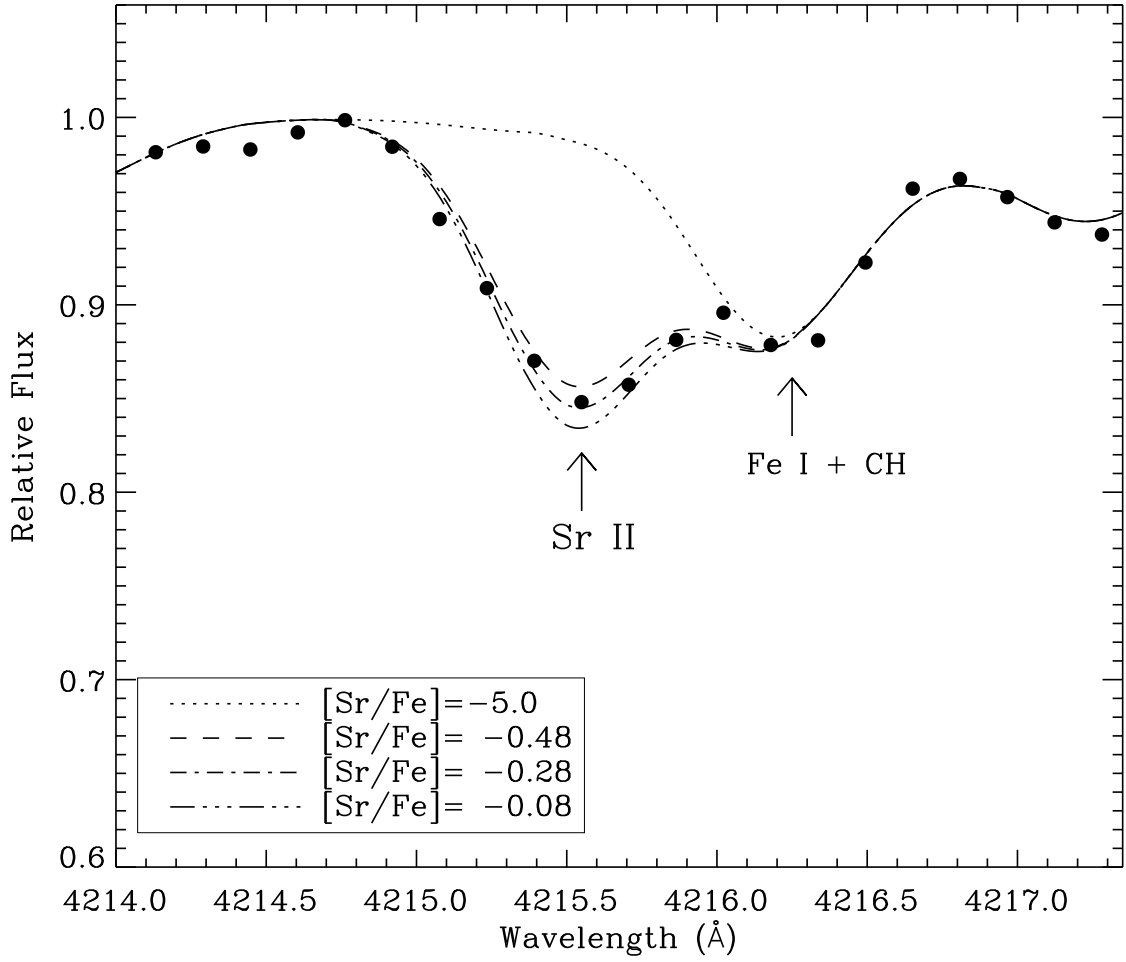}
\caption{Example of the synthesis for Sr II in the star
  CS 22957-022. 
The filled circles show the observed spectrum.\label{synthsr}}
\end{figure}

With the relatively low resolution of ESI, only the resonance lines of
Sr could be detected.  We derived the Sr abundance for our sample
via 
synthesis of the SrII line at 4215 {\AA}.  We used the log$gf$ value from
\citet{ivans06}, and the linelist from \citet{kurucz95} for
the surrounding lines in the region along with the LTE spectrum
 synthesis and
analysis program MOOG \citep{sneden73}, for the spectrum synthesis.
 Unfortunately the SrII 4077 {\AA} line proved
unsuitable because of blending with the 4077.3
{\AA} LaII line and 4078.0 {\AA} DyII line, both elements that we
cannot measure independently in our spectra.  Figure \ref{synthsr} shows
 a sample synthesis of the
4215 {\AA} line in the star CS 22957-022.  It is clear that the moderate
resolution of ESI makes spectral synthesis necessary to measure Sr
accurately. In particular we must ensure to take the Fe and CH blends
into account.

The carbon abundances were determined by spectral synthesis of the 4300
{\AA} region of the spectrum, which encompasses the
A$^2\triangle-$X$^2\Pi$ transitions of CH.  We used the linelist from
\citet{lucatello03} and MOOG for the spectral synthesis.

\begin{figure}
\plotone{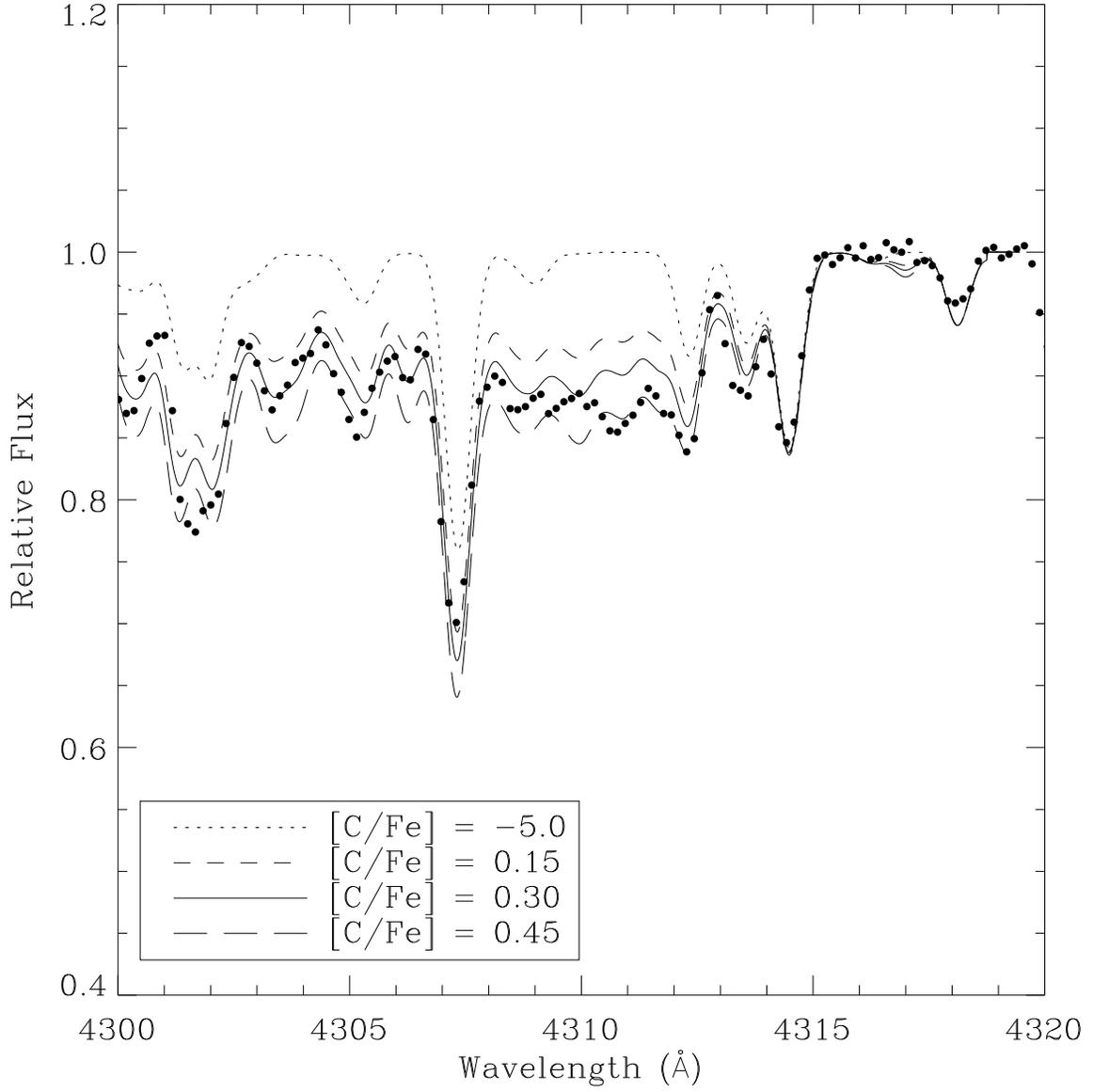}
\caption{Example of G-band synthesis for the star CS 22957-022. 
The filled circles show the observed spectrum.\label{synth}}
\end{figure}

Atmospheric parameters and models were taken directly from
Paper I.  We arbitrarily
assumed a value of 10 for the $^{12}$C/$^{13}$C ratios.  The
resolution of the spectra prevented us from performing 
measurements of this ratio.  The effects of this choice are discussed
$\S$ \ref{isotope}.  Figure \ref{synth} shows a sample of the synthesis. 

\begin{figure}
\plotone{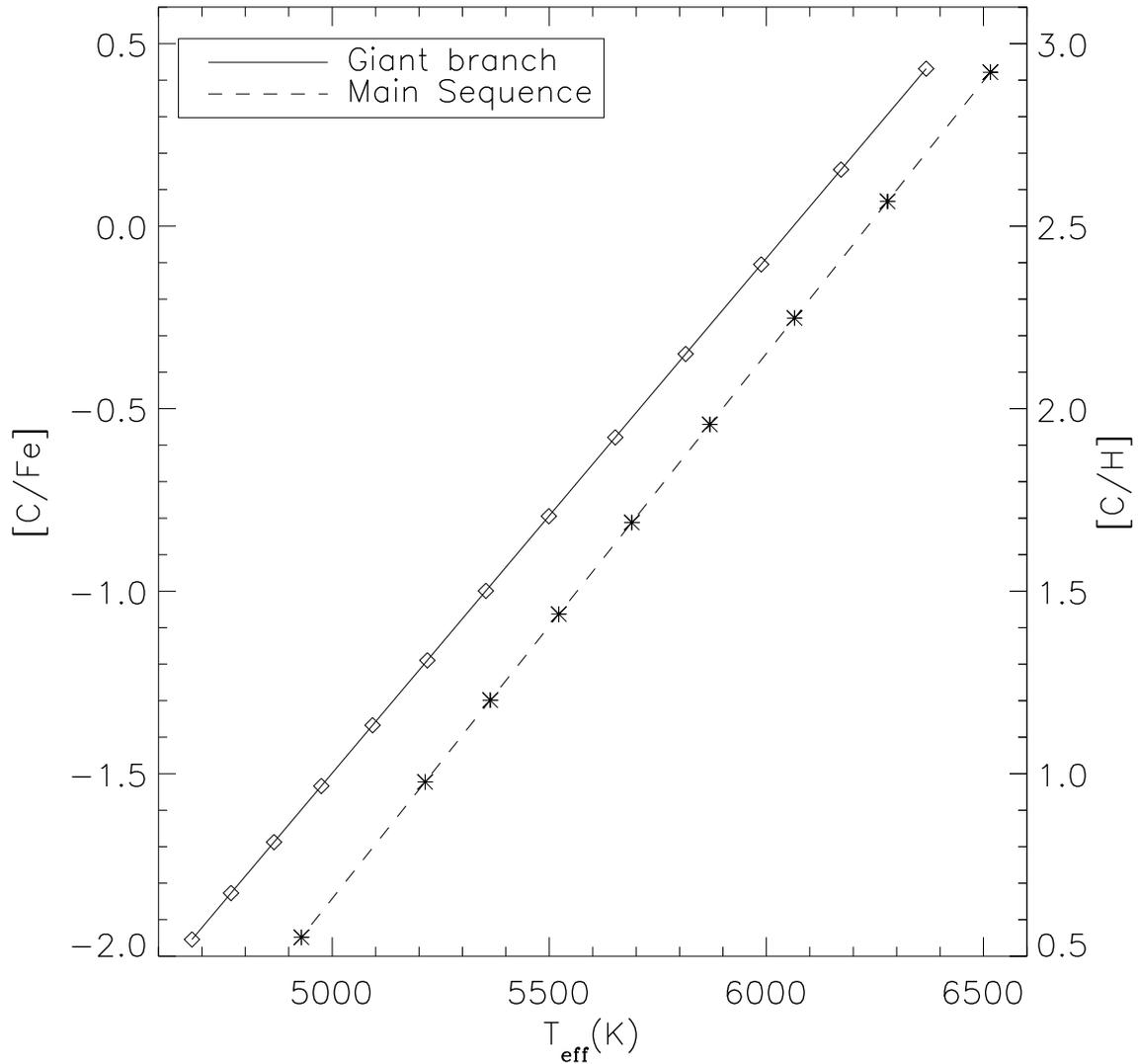}
\caption{Our estimated sensitivity limit to detecting [C/Fe] in
  our stars.  The value for [C/Fe] must be above the line at a given
  temperature to determine more than an upper limit. For increasing
  temperature, the log$g$s of the
  giants range from 1.45 to 3.87, and the log$g$s for the main
  sequence line range from 4.8 to 4.42. The plot shown
  is for a star with [Fe/H]=$-2.5$.\label{limit}}
\end{figure}

The relatively low resolution of ESI also affects the sensitivity we
have to measuring carbon abundances.  To test this effect we ran a set
of trial synthesis ranging from 4600 to 6500 K and with surface
gravities determined from the isochrones of \citet{bvand}.  The
metallicity of the models were all set to [Fe/H] of $-2.5$.  For our
range of temperatures, there are generally two possible choices for
the surface gravity, either along the main sequence or on the
giant/subgiant branch.  We calculated the lower limits of [C/Fe] that
we can measure for both cases and plot them in Figure \ref{limit}.

There are four main components to the carbon abundance error.
Uncertainties in the synthesis fit, uncertainties in the atmospheric
parameters, errors in the atomic line parameters, and, for C,  unknown
$^{12}$C/$^{13}$C ratios all contribute to the final abundance uncertainty. 

The uncertainty due to the synthesis fit was estimated by using the values
of the largest and smallest carbon abundance that could fit the data
by eye.  Figure \ref{synth} shows an example of these sample fits.
The typical fitting error was
between 0.1 to 0.2 dex.  

\subsubsection{Errors due to uncertainties in the adopted atmosphere
parameters}

We used three stars to represent the errors for objects in different
stages of evolution, BD+02 3375, CS 22183-031, and BS 16928-053.  We
then varied the atmospheric parameters, \teff, log$g$, [Fe/H], and
$\xi$, for each of these stars to determine the effect on the [C/Fe]
and [Sr/Fe] measurements. The results are presented in Tables
\ref{atmerrors} and \ref{sratmerrors}.  The results of varying [M/H]
in the atmospheres proved negligible in the final abundance
measurements.  We also took into account the \teff - log$g$ cross
term. Varying \teff by +125 K, log$g$ changes by $-0.03$, 0.15, and
0.29 dex for BD+02 3375, CS 22183-031, and BS 16928-053, respectively.

\subsubsection{$^{12}$C/$^{13}$C \label{isotope}}

The resolution of ESI prevented us from carrying out measurements of
the $^{12}$C/$^{13}$C ratio.  We instead used a value of 10 for the
entire sample when measuring the total carbon abundance.  This is
likely an incorrect assumption, so we chose the same three stars used
for the atmospheric error analysis to test the consequences of this
assumption.  We chose two extreme values of $^{12}$C/$^{13}$C, 4 and
200, and ran the synthesis again.  The effect was negligible for the
high case.  The [C/Fe] changed by less than +0.05 dex in all three
cases, a level that is not detectable in our spectra.  Using the lower
ratio value, $^{12}$C/$^{13}$C set to 4, the [C/Fe] changed by
$-0.05$, $-0.1$, and $-0.1$ dex for BD+02 3375, CS 22183-031, and BS
16928-053, respectively.

This comes about because the $^{13}$CH does not contribute
significantly to the absorption until it is a very high fraction of
the overall carbon abundance.  The effect is that we are only sensitve
to $^{12}$CH if we assume $^{12}$C/$^{13}$C greater than 10 for our
stars. Because of the resolution of our spectra, having a low
$^{12}$C/$^{13}$C does not noticeably change the shape of the
molecular feature, but it does change the depth. If the true value of
$^{12}$C/$^{13}$C$\simeq4$, then we have increased and hence
overestimated the overall carbon abundance to match the observations.
Therefore if the stars are heavily mixed compared to what was assumed
(very low $^{12}$C/$^{13}$C) then we may be overestimating [C/Fe] by
0.1 dex. This is only likely for the coolest giants in our sample,
\teff $\leq$ 4800 K, where deep mixing has likely taken place enough
to appreciably lower the $^{12}$C/$^{13}$C \citep{cayrel04}.

\subsubsection{Total uncertainty}

The total error for our [C/Fe] values is taken to be the quadrature
sum of the atmospheric and fitting error for each object.  The
contribution from uncertainty of the $^{12}$C/$^{13}$C ratio is not
included since we assume that it is negligible in most of our stars,
although it may be a systematic offset for our coolest giants. As shown in
Table \ref{abundances}, we find errors ranging from 0.17 to 0.26 dex.

The total error on our [Sr/Fe] measurements is taken to be the quadrature
sum of the atmospheric error for the stellar parameters closest to
each star, and 0.2 dex, the typical fitting error of the synthesis.
This corresponds to errors of 0.31,
0.32, and 0.39 dex for dwarfs, subgiants/turn-off stars, and giants,
respectively.  These relatively large errors are a consequence of
using a strong resonance line of SrII with its high sensitivity to the
choice of microturbulent velocity.

\subsection{Comparisons with Previous Studies}

\begin{figure}
\plotone{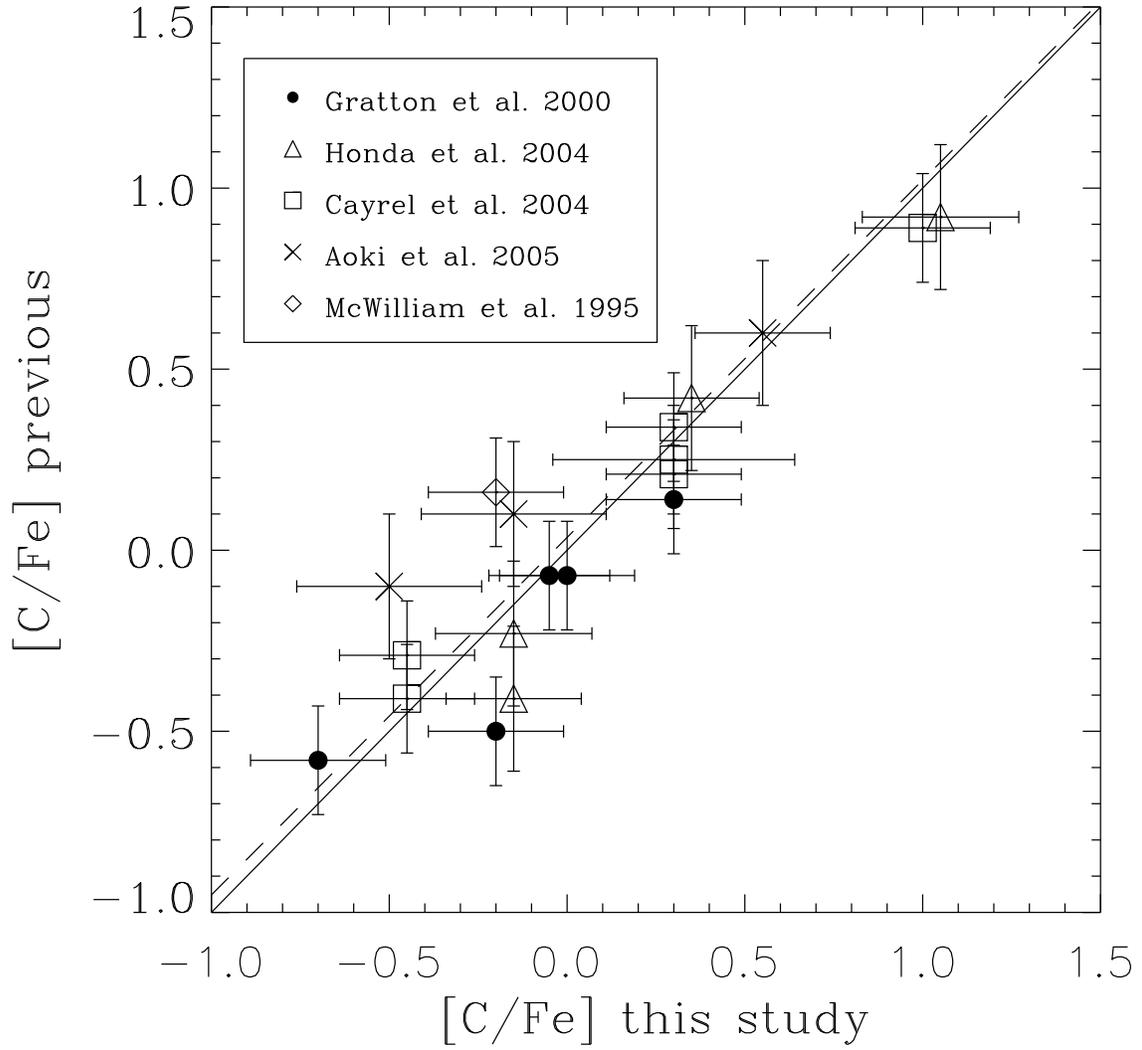}
\caption{A comparison of previous [C/Fe] measurements
 with this
  study. There is a slope of 0.99 to the best fit line (the dashed
  line)  The 1-1 line is shown as the solid line. 
 \label{vslit}}
\end{figure}

We compared our results for [C/Fe] with previously determined values
for 20 stars. We list these comparison values in Table
\ref{comparison} and plot them in Figure \ref{vslit}. The
dispersion between our study and the previous values is 0.21 dex, with
a mean offset of -0.03 dex in the sense of our values minus previous
values. This is consistent with our errors.

\begin{figure}
\plotone{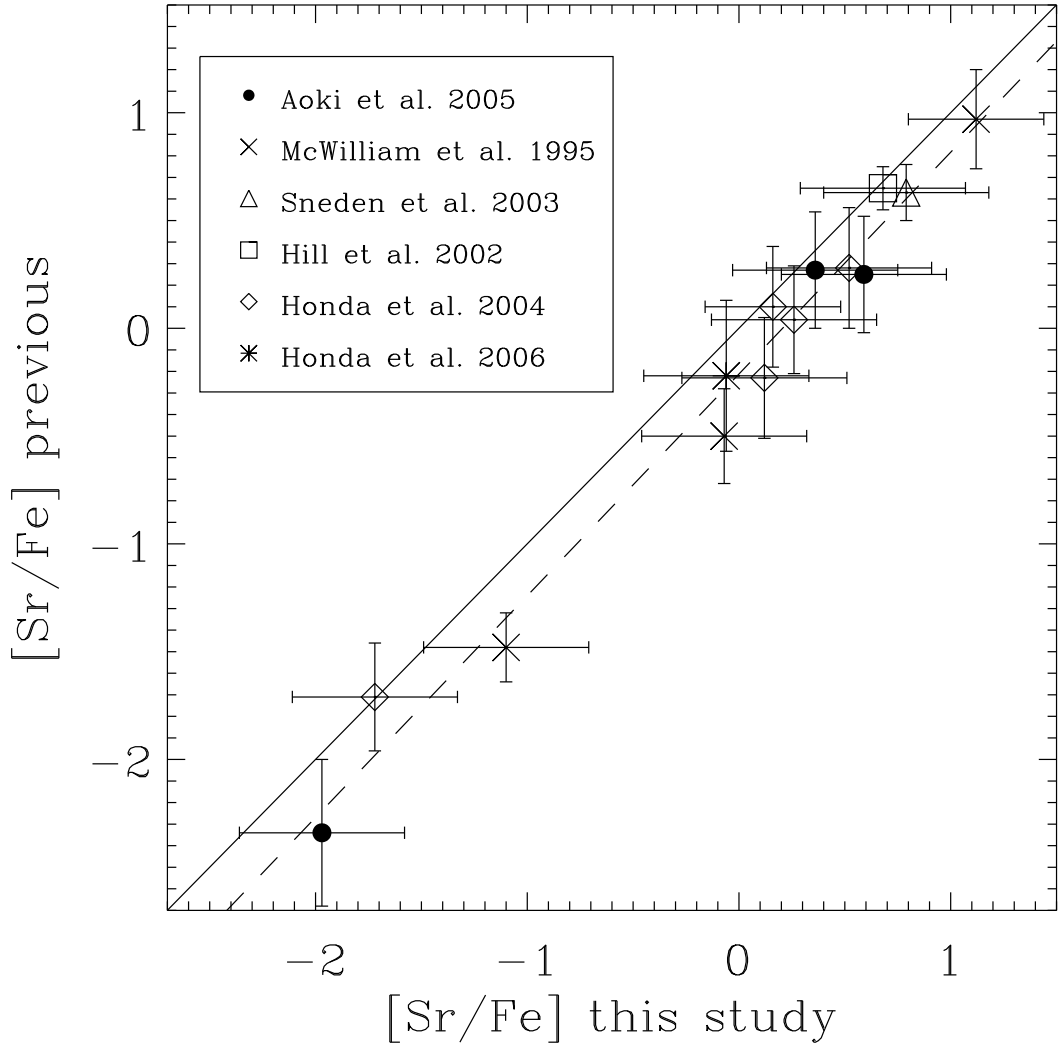}
\caption{A comparison of previous [Sr/Fe] measurements
  with this study.  The best fit line (the dashed line) has a slope of
  1.03.  
\label{compsr}}
\end{figure}

Table \ref{comparisonsr} lists our values for [Sr/Fe] and previously
measured values of [Sr/Fe].  From Figure \ref{compsr} we can see that
our [Sr/Fe] measurements are higher than previous studies.  Although
we have attempted to properly take into account all of the blends that
would affect our SrII line, this suggests that either we have
neglected to take into account a line/lines that do not affect the
high-resolution studies, or we are not treating the known blends
(e.g. the FeI line shown in Figure \ref{synthsr})
accurately. Also, comparing our stellar parameters with those from the
studies listed in Table \ref{comparisonsr}, we find that our \logg{}
values are on average 0.36 dex lower, while the \teff and microturbulent
velocities show no offset. As indicated in Table \ref{sratmerrors},
this means we may be deriving [Sr/Fe] values that are 0.05 to 0.1 dex
too high. The average offset between our measurements and those from the
earlier studies is 0.21 dex, a part of which that may be explained by
our \logg{} discrepancy.  However the scatter of the
$\Delta$[Sr/Fe] values is a relatively low 0.14 dex.  We are in
excellent agreement with previous measurements if we subtract off this
average offset.  It may be appropriate then, to lower our measured
[Sr/Fe] values by 0.21 dex when comparing our results with those from
high resolution studies.

There are three studies that were not included in the above
comparisons. Two of these included the stars HD 115444 and HD 122563.
\citet{jb2002}, measured [Sr/Fe] as $-0.27$ and $-0.39$ for HD 115444
and HD 122563, as compared to our measured values of 0.26 and
$-0.06$. For both these differences fall within the quoted errors
assuming the 0.2 dex offset, but only barely.  The main reason for
these differences is that the atmospheric parameters adopted for both
stars in \citet{jb2002} are quite different from those used in this
study.  Using their model atmosphere parameters and our measurements,
we find [Sr/Fe] values lowered to 0.06 and $-0.16$, for HD 115444 and
HD 122563, respectively.  Similarly, \citet{cohen04} find [Sr/Fe] =
0.06 for BS 16945-054 compared to our value of $-0.32$. However, if we
adopt their atmospheric parameters our measurement gives
[Sr/Fe]$=-0.07$.

The third study, by \citet{westin} gives [Sr/Fe] of 0.32 and 0.17 for
HD 115444 and HD 122563, respectively. Both of these values are
greater than our measurements, which is at odds with the other
comparison results summarized in Table \ref{comparisonsr}. In addition
the \citet{westin} HD 122563 and HD 115444 [Sr/Fe] values are much
higher than the abundances from \citet{honda06} and
\citet{honda04}. Unlike the case for \citet{jb2002}, the atmospheric
parameters for both stars from \citet{westin}, the studies just
listed, and ours study are in reasonable agreement, so it is unclear
why the \citet{westin} abundances are different. Although we have
chosen to use the \citet{honda06} and \citet{honda04} values for our
comparisons above, it should be noted that the Westin measurements
taken by themselves are entirely consistent with our abundances. If we
did adopt the Westin values, we would find an overall offset of
0.16 dex from previous studies, and a scatter of 0.20 dex in
$\Delta$[Sr/Fe].

\section{Results \label{results}}

Table \ref{abundances} summarizes our values for [Sr/Fe] and [C/Fe]
relative to the solar abundances of \citet{anders89}, along with the
fitting, atmosphere, and final errors for [C/Fe].  Figure \ref{trends}
shows [C/Fe] plotted against [Fe/H], \teff, and log$g$.  We see a
trend of decreasing [C/Fe] with decreasing \teff beginning at 5000 K,
which is also mirrored by the log$g$ plot.  This is consistent with
the results of \citet{cayrel04} where in giants deep mixing has
brought material to the surface from layers where carbon has been
converted to nitrogen \citep{gratton00}.  Unfortunately we cannot
obtain nitrogen abundances from our spectra to verify this in our
stars, as was done by \citet{spite05} for the Cayrel sample.

There is a large scatter over the entire sample with a dispersion of
0.52 dex.  The average for [C/Fe] over the entire sample is 0.11 dex.
Below [Fe/H]$ = -2.0$ we note that the scatter is 0.59 dex, compared
to only 0.33 dex for more metal-rich objects.  This reflects the fact
that all extremely high [C/Fe] values are found for
[Fe/H]$<-2.0$. Considering only stars with [Fe/H]$<-2.0$, we find that
5 out of 53 stars, or 9$\pm4$\% (assuming a bimodal distribution), can
be classified as CEMP stars among the entire sample. If we confine the
sample to stars with \teff $> 5200$K, which we argue below represent a
sample unbiased in C-richness, we find 3 out of 26 stars, or
12$\pm6$\% fall into the CEMP category.

\begin{figure}
\plotone{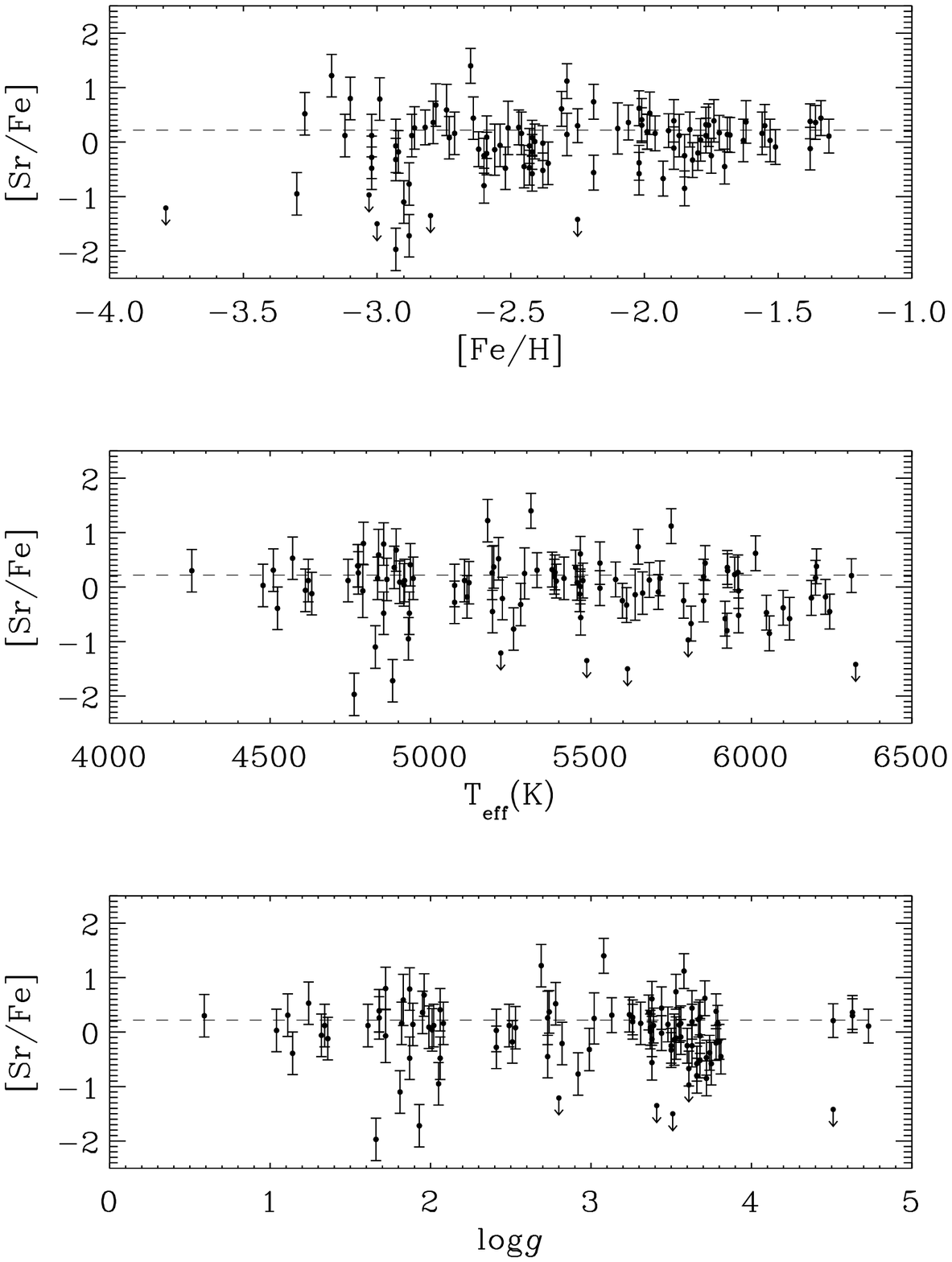}
\caption{Plots of [Sr/Fe] versus [Fe/H],
  \teff, and log$g$.\label{srtrends}}
\end{figure}

Figure \ref{srtrends} shows how [Sr/Fe] varies over [Fe/H], \teff{},
and \logg{}.  There are no obvious trends with \teff{} or
\logg{}.  As has been seen in previous studies
(e.g. \cite{mcwilliam95}, there does appear a marked increase in the
scatter of [Sr/Fe] below [Fe/H]$\simeq-2.3$, as well as an increase in
the number of very low [Sr/Fe] abundance measurements.  This reflects
the inhomogeneity of the early ISM and the stochastic nature of the
neutron-capture process that produced Sr.

\section{Discussion}

\subsection{Sr in the early Galaxy}

\begin{figure}
\plotone{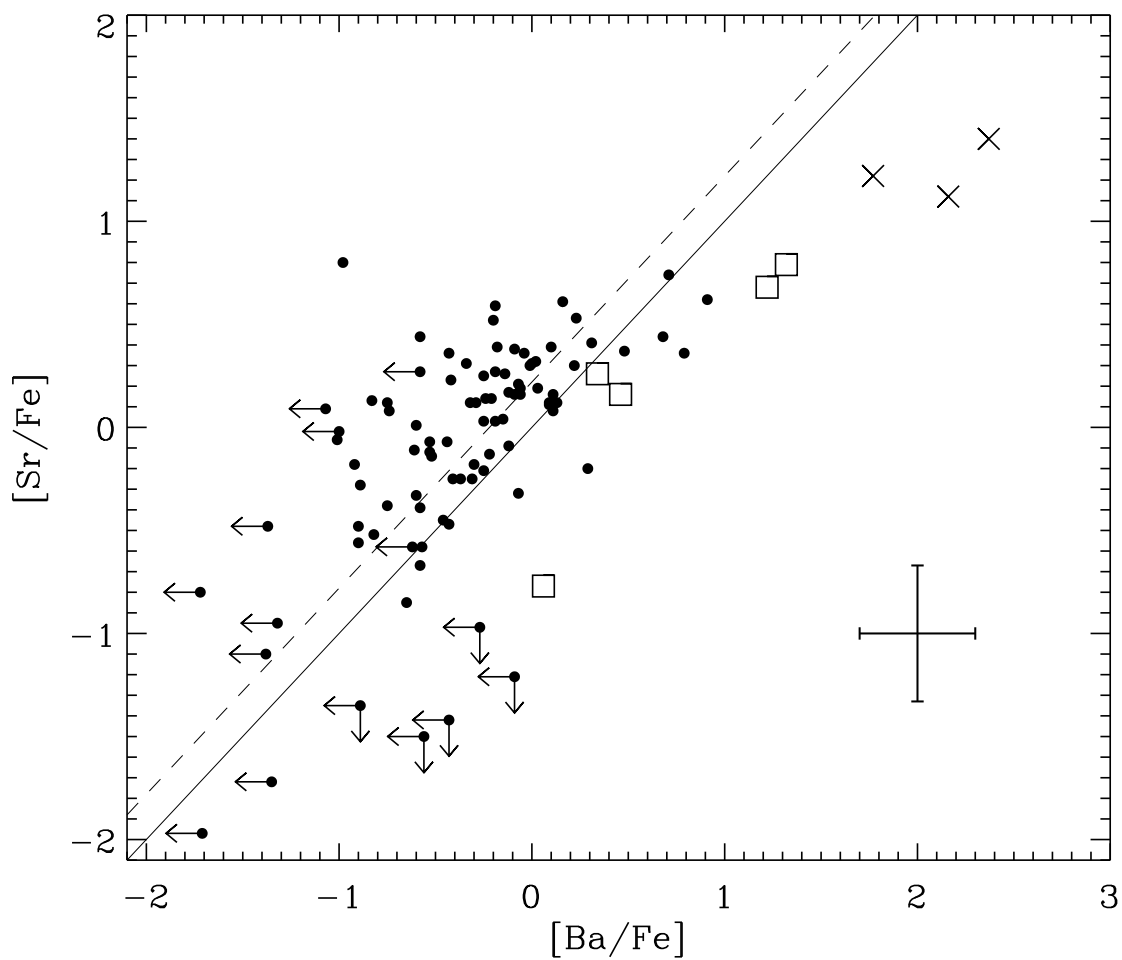}
\caption{A plot of [Ba/Fe] vs. [Sr/Fe].  The one-to-one value
  is shown as the solid line.  The dashed line is the one-to-one line
  offset by 0.23 dex.  A representative error bar is shown in the
  lower right.  Known r-process enhanced stars are plotted with a
  square symbol, and known s-process enhanced stars are plotted with
  the `x' symbol.  They are HD 115444, BS 16981-009, CS 22183-015, CS
  22183-031, CS 22892-052, CS 22898-027, CS 31062-050, and CS
  31082-001. 
\label{srvsba}}
\end{figure}

\begin{figure}
\plotone{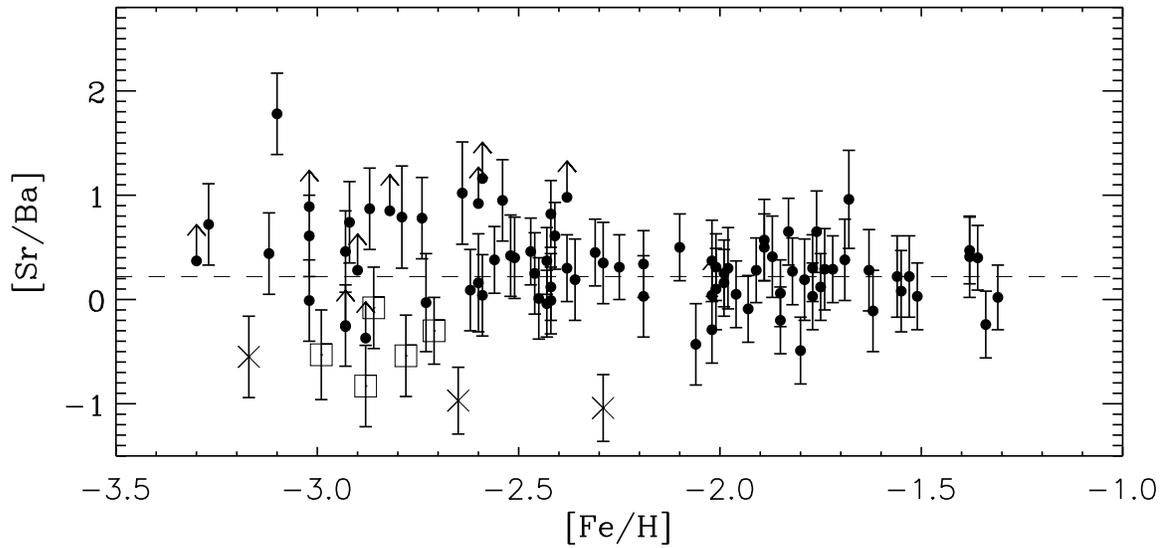}
\caption{Here we plot [Sr/Ba] as it varies over our range of
  [Fe/H].  It is a clear offset above zero for the average of [Sr/Ba].
  The scatter also noticeably increases below [Fe/H]~$-2.0$. The
  symbols are the same as in the previous figure. 
\label{srba}}
\end{figure}

Combining our Sr abundances with Ba abundances from Paper I allows us
to see the evidence of different neutron-capture events in an
individual star.  We show this in Figures \ref{srvsba} and \ref{srba}.
The three stars with the highest values of [Ba/Fe], CS 22183-015
\citep{jbcs22183}, CS 31062-050 \citep{jbcs31062}, and CS22898-027
\citep{mcwilliam95}, have been shown to be heavily polluted by
s-process material. The square symbols mark the stars known to have an
r-process signature; HD 115444 \citep{westin}, BS 16981-009 (also
called HE1430+0053, \citealt{heres}), CS 22183-031 \citep{honda04}, CS
22892-052 \citep{sneden96,sneden03}, and CS31082-001
\citep{hill02}. Six of eight of these stars appear separated from the
majority of the points in Figures \ref{srvsba} and \ref{srba}.  The
two exceptions are HD 115444 and CS 22183-031.  They have neither a
high [Ba/Fe], nor particularly low [Sr/Ba]. Using lower resolution
spectra, we can identify highly r/s-process enhanced stars, but we
would not be complete in doing so.

The scatter of our [Sr/Ba] values at [Fe/H]$>-2.2$ is 0.29, well
within our errors. At [Fe/H]$\leq-2.2$, however, the scatter is 0.55
dex.  This indicates that multiple processes are contributing to the
light neutron-capture synthesis.  As is clearly seen in figure
\ref{srvsba} many of our stars are measured at or above [Sr/Ba]=0, far
from where the main s- and main r-process enhanced stars sit.  This
demonstrates that in the early galaxy Sr was being produced in some
other process, whether it was a weak s- or r-process \citep{jb2002,
  kratz07}, or some other unidentified process \citep{travaglio}.  In
particular EMP stars such as BS 16550-087, with a [Sr/Ba] of 1.78, may
prove ideal for testing different scenarios for Sr production.

\subsection{Classification of the C-rich stars}
\begin{figure}
\plotone{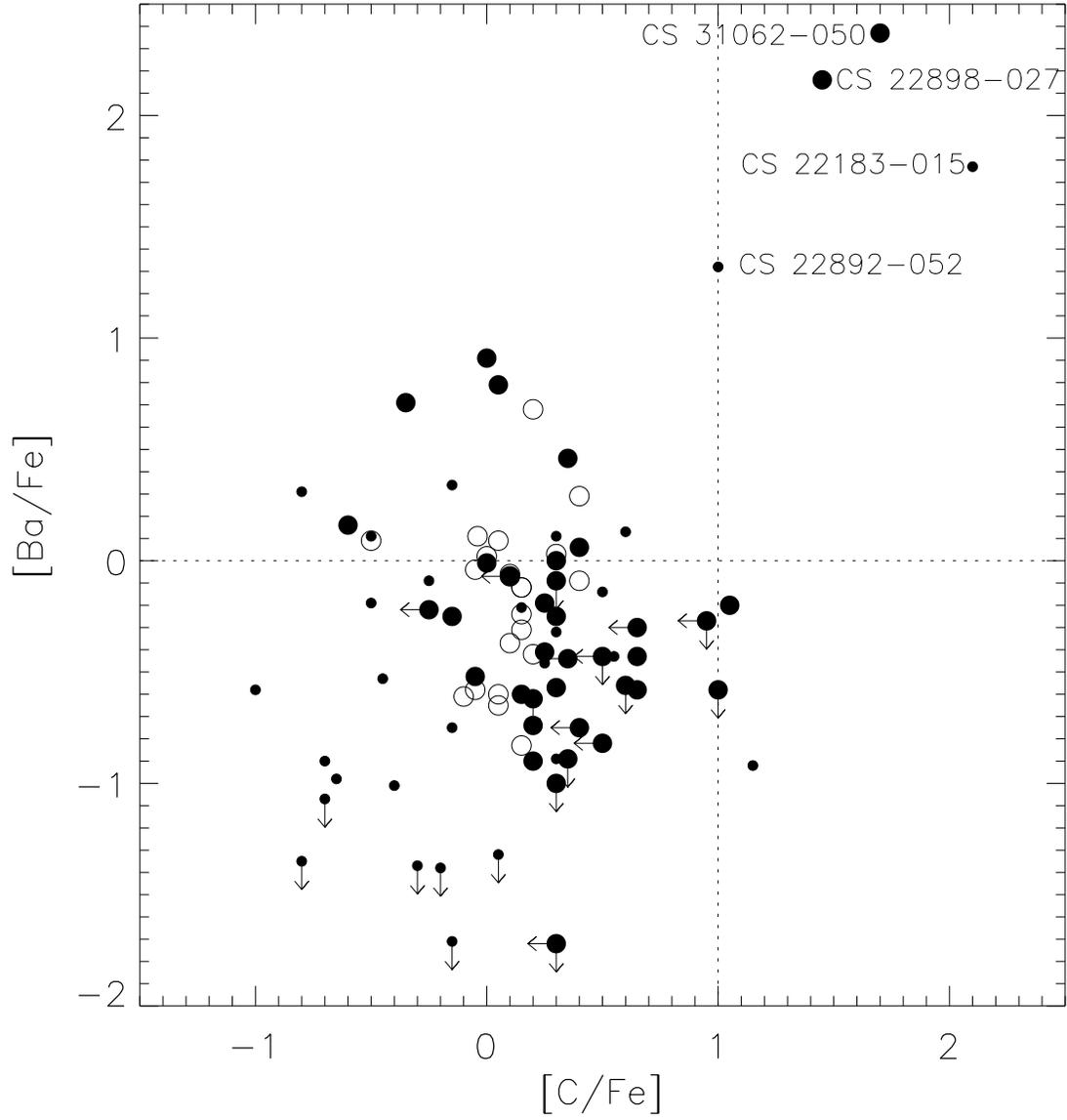}
\caption{Plot of [Ba/Fe] versus [C/Fe].  The filled
  circles represent stars with [Fe/H] $<-$2.0.  The large circles
  represent stars with T$_{eff}\geq 5200$K, and the small circles stars
  with T$_{eff}<5200$K.
  \label{ba}}
\end{figure}

The Ba abundances also allow us to place the CEMP stars into
subclasses.  We plot [Ba/Fe], derived in Paper I, versus [C/Fe] in
figure \ref{ba}.  The three stars in the upper right hand corner of
the plot all have follow-up data from which Eu could be measured,
revealing that CS 22183-015, CS 31062-050, and CS 22898-027
\citep{aoki02c} are CEMP-r/s stars.

The object CS 22892-052 has also been labeled in figure \ref{ba}.  As
detailed by \citet{sneden96,sneden03}, this is a very r-process-rich
object that is also carbon-rich (CEMP-r).  With a few exceptions CS
22892-052 fits the scaled solar system r-process abundance extremely
well, suggesting that its barium abundance has a predominately
r-process origin.  This has been the only CEMP and r-process-rich star
yet discovered \citep{ryan05}, and it suggests that the carbon
over-abundance is not connected to it's high barium abundance.  This
is also a way to differentiate r and s-process stars if a high [Ba/Fe]
is measured, even when a r-process element such as europium is not
measured.  As all other r-process enhanced EMP stars, such as CS
31082-001, are not C-rich then a high [Ba/Fe] with normal [C/Fe] is
an indication of r-process and not s-process enhancement.

Three stars in figure \ref{ba} have [C/Fe]$\geq$1, but [Ba/Fe]$<0$: BS
16077-007 with [Fe/H]$=-2.8$, BS 16929-005 with [Fe/H]=$-3.3$ and CS
29502-092 with [Fe/H]$=-2.9$. The lack of Ba
indicates that these stars have not been polluted by a large s-process (or
the r/s process) event, and a different mechanism must have taken place to
explain their C-enhancement. An extreme example of this type of object
is CS 22957-027, analyzed by \citet{nrb97}, with [C/Fe]
$= 2.2$ and [Ba/Fe] $= -1.0$.  
According to Paper I, BS 16077-007
has [Mg/Fe]$=0.21$, BS 16929-005 has [Mg/Fe] $=0.63$, and CS 29502-092
has [Mg/Fe] $=0.42$ dex. BS 16929-005 has also been observed by
\citet{honda04}, who measured a [Mg/Fe] of 0.38 dex.  
CS 29502-092 was also observed by \citet{aokicempno} with much
higher resolution, giving results consistent with ours. Therefore, these
stars are not enriched in the $\alpha$ elements as well as C, and
are not members of the same class as CS 29498-043, which has high
[C/Fe], low [Ba/Fe], but a [Mg/Fe]$=1.81$ \citep{aokialpha}. One
possibility is that these objects are the result of stellar formation
from carbon-rich gas. Or, as suggested by \citet{nrb97}, it
could come from a mechanism described by \citet{fujimoto00} where 
a helium core flash in a zero metal star induces hydrogen
burning, and the subsequent material is then dredged up to the surface
layers.

Overall, our sample seems to fall into three classes.  Using
the designation of \citet{ryan05}, there are two classes of
carbon-rich metal-poor stars, one that is rich in the s-process and
one that is normal in the s-process.  Then there are the barium and
carbon `normal' objects that take up the majority of Figure \ref{ba}.
These include many objects that are labeled evolved giants.  As
described in \cite{lucatello06}, it is possible that their carbon
abundance were higher in the
past, but due to dilution by mixing processes are lower now.  

\subsection{Fraction of C-rich stars}

\begin{figure}
\plotone{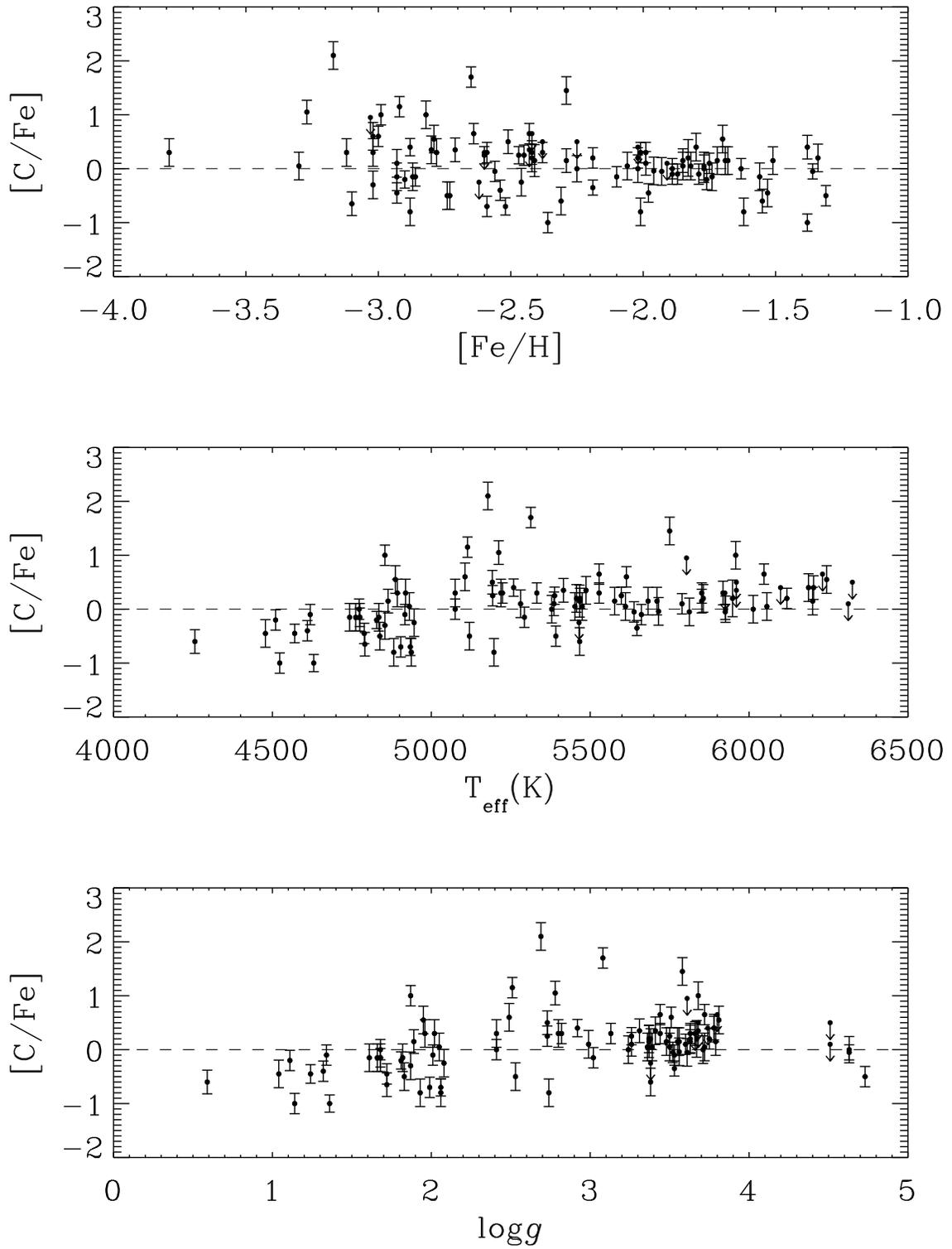}
\caption{Plots of [C/Fe] versus [Fe/H],
  \teff, and log$g$.\label{trends}}
\end{figure}

One of the most helpful ways for understanding how C-rich stars were
formed is determining their relative fraction at different
metallicities. Aggregate percentages of C-rich stars ([C/Fe]$\geq
1.0$) with [Fe/H] values less than a particular metallicity have been
reported by a number of authors. As discussed in the Introduction,
these estimates range from 25\% to 9\%. Several reasons for the
discrepancies have been identified. \cite{cohen05} showed that the CaK
and H$\delta$ indices that have been used in combination to derive
metallicities by BPS, the HES \citet{hes} and the INT study
\citet{int}, among other others, have C molecular features in the
``continuum'' sidebands. The effect on the H$\delta$ feature suggests
that the star is cooler than it actually is, which translates into a
lower [Fe/H].  The CaK feature also appears to be weaker, which
further increases the discrepancy between the measured and the real
[Fe/H].  The [C/Fe] ratios in these stars are then artificially
high. A second effect is the real decline in C abundance as stars move
up the red giant branch \citep{lucatello06}.  Both effects can be
ignored for stars with \teff $> 5200$ where the CH lines around the
CaK feature are insignificant \citep{cohen05} and the depletion of C
in red giants has not begun (see Figure \ref{trends}). Of course, this
reduces the sample of stars available for calculating the fraction of
C-rich stars, and makes the addition of more metal-poor stars with
complete abundances important. 

Possible reasons for this study to be biased toward or against finding
C-rich stars are the effects of C-richness on metallicity estimators
used in this study. In particular, if C-richness makes the star appear
more iron-poor than it actually is, then it would have a greater
chance of showing up in our study. For the sources for our study, the
$ubvy$ photometry of \cite{twarog00} is unaffected by molecular carbon
features \citep{schuster04}, The INT study \citep{int} used the CaK,
CaI line at 4226\AA{} and the hydrogen with a neural network to
calculate metallicities. As with BPS, this technique should be
relatively unaffected for stars with \teff$>$5200K. Finally, we used
the UBV photometry of \citet{norris99} to identify metal-poor
candidates by their UV excesses. To test whether the UBV colors were
sensitive to C abundance as well as metallicity, we found UBV
photometry for as many CEMP stars as we could, as well as confirmed
non-CEMP stars of similar metallicities in the photometric catalogs of
\citet{preston91}, \citet{norris99}, and \citet{beers07}.  If we
confine ourselves again to stars with higher temperatures, we found
that very metal-poor carbon-rich stars and carbon-normal metal-poor
stars fall in the same part of the UBV diagram, so we expect our
selection based on UV excess to be essentially unaffected by stars of
differing [C/Fe] . The sample size of very metal-poor stars with
complete UBV photometry and abundances is small, but the photometric
result is consistent with the molecular bands being weaker in the
bluer stars. We therefore conclude that our sample of stars with
\teff$>5200$K is only negligible affected by bias toward finding
carbon-rich stars.

\begin{figure}
\plotone{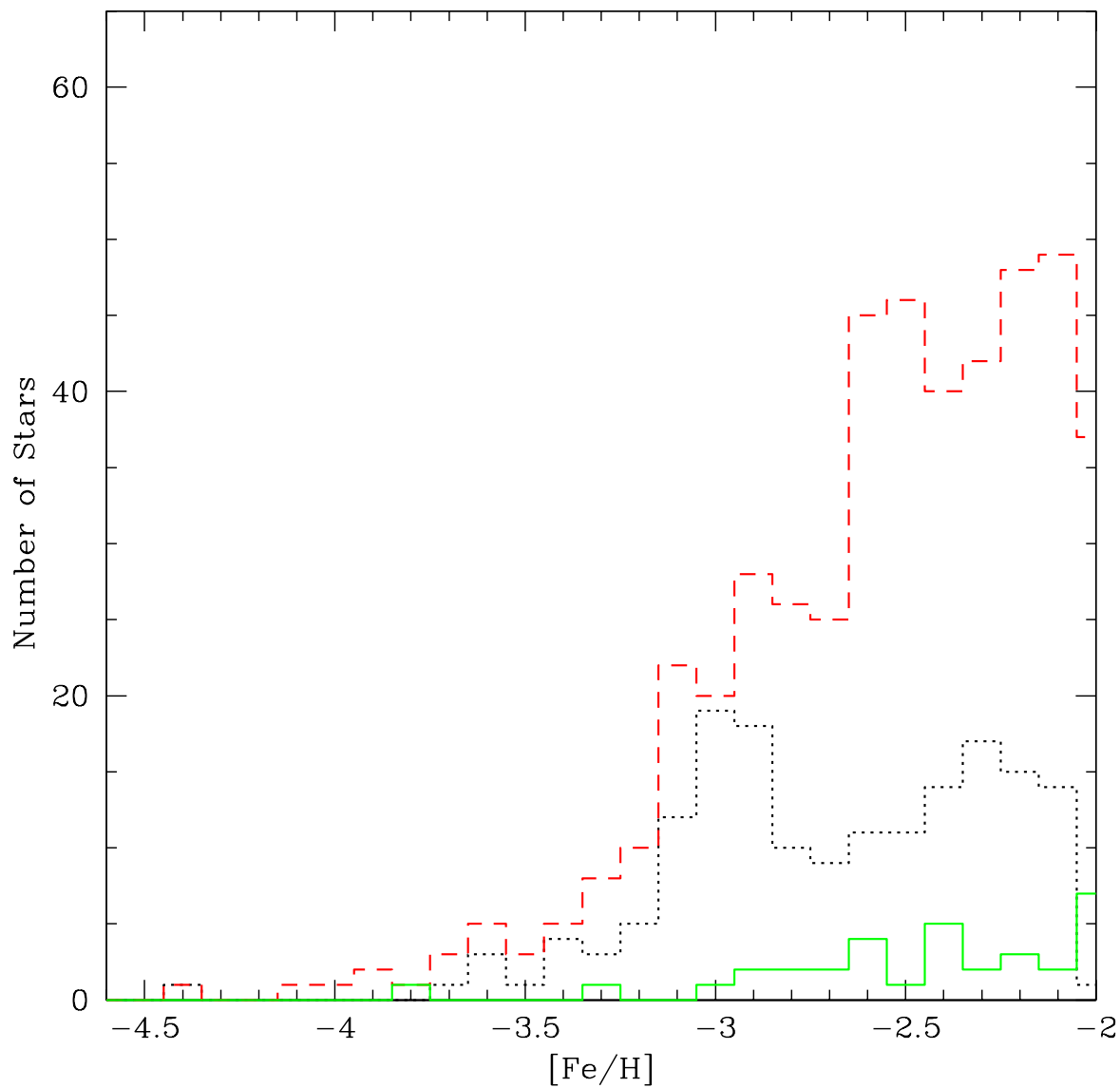}
\caption{The metallicity
distribution function for BPS (dashed line), \citet{frebel06} (dotted
line) and this study (solid line) for all stars with
[Fe/H]$<-2.0$. For this study, the sample was also restricted to stars
with \teff$>$5200K.  The other samples had C-rich fractions quoted for
the entire sample, so they were not restricted in \teff. See the
electronic edition of the Journal for a color version of this figure.
\label{mdf}}
\end{figure}

There is one additional effect that must be mentioned. There is
increasing observational evidence that the fraction of C-rich stars
climbs as [Fe/H] decreases. Given the apparent dependence of the
fraction of stars that are C-rich on [Fe/H], a survey that was
particularly efficient at finding stars with [Fe/H] $<-4$ would report
a higher C-fraction than one that was not, even if both groups overall
studied stars with [Fe/H]$<-2.0$. In Figure \ref{mdf}, we plot the
metallicity distribution functions at [Fe/H]$<-2.0$ for BPS,
\citet{frebel06} and this study. There are clear differences in the
MDFs. The BPS and our MDF in this metallicity range peak at
[Fe/H]$\simeq-2.0$, while the \citet{frebel06} MDF has two peaks, the
strongest at [Fe/H]$\simeq-3.0$ and a lower one at [Fe/H]$\simeq-2.3$. Also,
although our MDF has the same peak range as that of BPS, we do not
sample the space below [Fe/H]$=-3.2$ well, a result of our much
smaller sample size. While some of this is likely caused by different
metallicity scales, arising, for example, from different temperature
scales, that does not explain all the differences in the observed
MDFs. These differences will be reflected in the percentage of C-rich
stars.

We have directly tested for the effect of differing observational MDFs
on derived C-rich fractions. Using the MDFs shown in Figure \ref{mdf},
we have recalculated the C-rich percentage if our sample had the same
MDF as \citet{frebel06} (because of the small number of stars with
[Fe/H] $<-3$ in our sample, in this example we put all of the stars below this
metallicity into the [Fe/H]$-3.0$ bin for both MDFs). This was done by
first finding the percentage of C-rich stars in each of the bins (0.1
dex in [Fe/H]) for our sample.  We then applied this percentage to the
each respective bin of the Frebel MDF, and recalculated the percentage
of C-rich stars with [Fe/H]$<-2.0$. We find the fraction of C-rich
stars to increase by 3\%, a small but noticeable amount. We also
carried this out using the BPS MDF in place of the Frebel MDF, and we
find that the C-rich percentage increases by 2\%. The reason for this
becomes clear when examining Figure \ref{mdf}. Both the Frebel and BPS
MDFs have a higher percentage of their stars below [Fe/H]$=-2.75$
compared to ours, and having more stars at lower metallicity will
increase the overall C-rich fraction.

\begin{figure}
\plotone{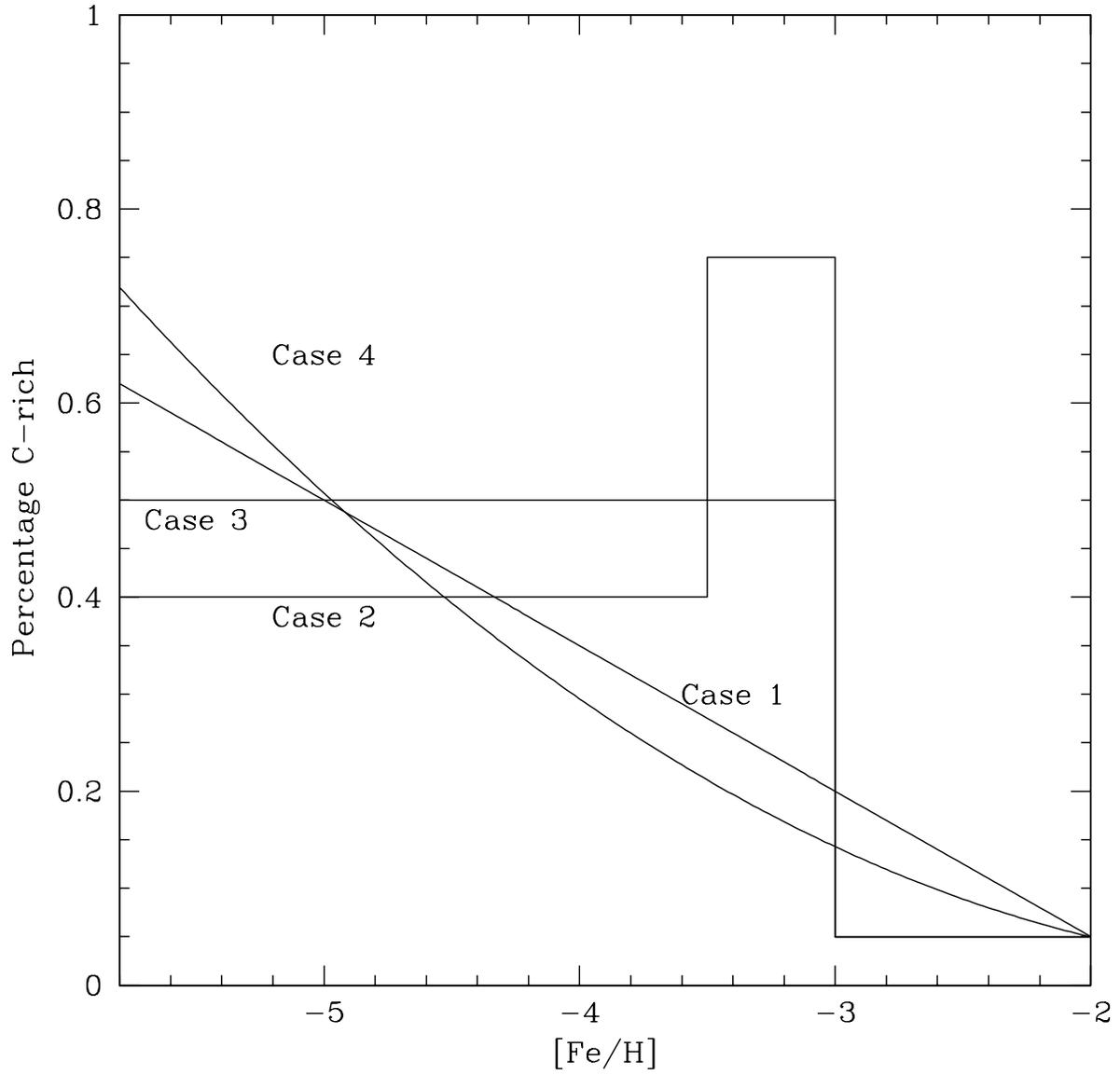}
\caption{Four examples of possible
distributions of C-rich stars vs. [Fe/H]. 
\label{cdist}}
\end{figure}

To further provide some estimate of the size of the effect, we have
selected four possible distributions of the fraction of C-rich objects
as a function of [Fe/H] for [Fe/H]$<-2.0$. These are illustrated in
Figure \ref{cdist}. The distributions are described below, with
$f$(CEMP) standing for the fraction of stars with [C/Fe]$\geq 1.0$.
\begin{eqnarray*}
\mbox{Case 1:} \quad f(\mbox{CEMP}) &=& -0.15\times \mbox{[Fe/H]}-0.25 \\
\mbox{Case 2:} \quad f(\mbox{CEMP}) &=& 0.05 \qquad \quad \mbox{[Fe/H]} > -3.0\\
  &=& 0.80 \qquad \quad -3.0 > \mbox{[Fe/H]} > -3.2\\
  &=& 0.40 \qquad \quad \mbox{[Fe/H]}<-3.2\\
\mbox{Case 3:} \quad f(\mbox{CEMP}) &=& 0.05 \qquad \quad \mbox{[Fe/H]}>-3.0\\
  &=& 0.50 \qquad \quad \mbox{[Fe/H]}<-3.0\\
\mbox{Case 4:} \quad f(\mbox{CEMP}) &=& 0.04+0.06\times\mbox{[Fe/H]}+0.030\times\mbox{[Fe/H]}^2
\end{eqnarray*}

These cases were mainly motivated by having a lower fraction of CEMP
stars at higher metallicity than at lower metallicity. In Case 2 and
Case 3, the transition is a sharp jump, while in Case 1 and Case 4,
the transition is more gradual. Case 2 includes the possibility that a
large number of CEMP-s stars are created at that particular
metallicity, if it is possible for a large fraction of lower-mass AGB
stars to be created at that metallicity. These are by no means the
only possibilities and are merely meant to illustrate the range of
CEMP fractions that can occur.  The expected fraction of C-rich stars
for each case, given the MDF of a particular study are listed in Table
\ref{cfedist}. The errors were calculated from the standard error of
the sample for 1000 Monte Carlo realizations of drawing stars from the
given MDFs.

It is clear that including only stars with [Fe/H] $<-2.5$ results in
higher fractions than if a more metal-rich ([Fe/H]$< -2.0$) cut is
used, and comparisons between studies should at the very least be done
using the same metallicity cut. Table \ref{cfedist} also illustrates
that even if the same metallicity range is used, very different
fractions of C-rich stars can be observed for the same underlying
distribution.  In Case 2, the range is 11\% to 29\%, an effect that is
entirely due to the different MDFs for stars more metal-poor than
[Fe/H] $< -2.5$.  Therefore this study's reported fraction of 12\%
C-rich stars for [Fe/H]$<-2.0$ (admittedly with large uncertainties)
is not incompatible with the higher fractions reported by
\citet{marsteller05} for the BPS sample for [Fe/H]$<-2.5$. The
fraction of 9\% from \citet{frebel06} is not duplicated by the current
choices for $f$(CEMP). However, because it is also probable that if
the fraction of C-rich stars increases with distance from the plane,
as reported by \citet{frebel06}, then the focus of that paper on
brighter stars and the focus of this paper on fainter ones, could
provide an additional explanation of the offset since we may then
expect the $f$(CEMP) to have a different form. Therefore, continued
work on the fraction of C-rich stars must either adequately sample the
MDF or be corrected for its effects to determine an accurate form for
the $f$(CEMP), and also must take into account the possible effect of
location relative to the galactic plane. Larger sample of stars, for
example from the SDSS survey \citep{marstellerAAS}, show promise in
providing the most bias-free results in this area.

\acknowledgements
This research is based in part upon work supported by the 
National Science Foundation under Grant Number AST-0098617
Any opinions, findings, and conclusions or recommendations 
expressed in this material are those of the author(s) and do not necessarily
reflect the views of the National Science Foundation.
The authors wish to recognize and
acknowledge the very significant cultural role and reverence that
the summit of Mauna Kea has always had within the indigenous Hawaiian
community.  We are most fortunate to have the opportunity to conduct
observations from this mountain.

\bibliography{ms}

\clearpage

\begin{deluxetable}{lccccccc}
\tablecolumns{8}
\tablewidth{0pc}
\tablecaption{Sensitivity of Carbon abundance to Atmospheric
  Parameters, where the relative change in Carbon is given in dex. \label{atmerrors}}
\tablehead{
\colhead{Star} & \colhead{\teff} & \colhead{log$g$} & \colhead{$\xi$} & \colhead{$\Delta$[C/Fe]}  
& \colhead{$\Delta$[C/Fe]} & \colhead{$\Delta$[C/Fe]} & \colhead{}
\\ \colhead{ID} & \colhead{(K)} & \colhead{(dex)} & \colhead{(km s$^{-1}$)} &\colhead{\teff +125
  K} & \colhead{log$g$ +0.5 dex} & \colhead{$\xi$ +0.5 km s$^{-1}$} & \colhead{Total}}
\startdata
BD+02 3375\tablenotemark{*} & 5926 & 4.63 & 1.0 &  0.08  &  0.09  &  0.07  & 0.14 \\
CS 22183-031                 & 5416 & 3.31 & 1.5 &  0.10  & $-$0.06  &  0.12  & 0.16 \\
BS 16928-053                 & 4743 & 1.61 & 2.0 &  0.12  & $-$0.09  &  0.13  & 0.16 
\enddata
\tablenotetext{*}{Log$g$ is varied by $-$0.5 dex for this object because
of its already high initial value.}
\end{deluxetable}

\clearpage

\begin{deluxetable}{lccccccc}
\tablecolumns{8}
\tablecaption{Sensitivity of Strontium abundance to Atmospheric
  Parameters, where the relative change in Strontium is given in dex. \label{sratmerrors}}
\tablehead{
\colhead{Star} & \colhead{\teff} & \colhead{log$g$} & \colhead{$\xi$} & \colhead{$\Delta$[Sr/Fe]}  
& \colhead{$\Delta$[Sr/Fe]} & \colhead{$\Delta$[Sr/Fe]} & \colhead{}
\\ \colhead{ID} & \colhead{(K)} & \colhead{(dex)} & \colhead{(km s$^{-1}$)} &\colhead{\teff +125
  K} & \colhead{log$g$ +0.5 dex} & \colhead{$\xi$ +0.5 km s$^{-1}$} & \colhead{Total}}
\startdata
BD+02 3375\tablenotemark{*} & 5926 & 4.63 & 1.0 &  0.04  &  $-$0.10  &  $-$0.22  & 0.24 \\
CS 22183-031                 & 5416 & 3.31 & 1.5 &  $-$0.06  &  0.09  &  $-$0.23  & 0.25 \\
BS 16928-053                 & 4743 & 1.61 & 2.0 &  $-$0.13  &  0.06  &  $-$0.32  & 0.34 
\enddata
\tablenotetext{*}{Log$g$ is varied by $-$0.5 dex for this object because
of its already high initial value.}
\end{deluxetable}

\clearpage

\begin{deluxetable}{lrrrrrrccc}
\tablecolumns{10}
\tablewidth{0pc}
\tabletypesize{\scriptsize}
\tablecaption{Atmosheric parameters and abundances \label{abundances}}
\tablehead{
\colhead{Star} & \colhead{T$_{eff}$}   & \colhead{log$g$} & \colhead{[Fe/H]}
&\colhead{[Sr/Fe]} &\colhead{[Ba/Fe]}  & \colhead{[C/Fe]} &\colhead{Synthesis} &
\colhead{Atmosphere} & \colhead{Total} \\ \colhead{ID} & \colhead{(K)} &
\colhead{(dex)} & \colhead{(dex)} & \colhead{(dex)} & \colhead{(dex)} & \colhead{dex} &
\colhead{[C/Fe] error} & \colhead{[C/Fe] error} & \colhead{[C/Fe] error}}
\startdata
  BD-03 2525 & 5789 &  3.60 & $-$1.75 & $-$0.25 & $-$0.37 &  0.10 &  0.10 &  0.16 &  0.19  \\
  BD+02 3375 & 5926 &  4.63 & $-$2.25 &  0.30 & $-$0.01 &  0.00 &  0.20 &  0.14 &  0.24  \\
  BD+23 3130 & 5224 &  2.82 & $-$2.59 & $-$0.21 & $-$0.25 &  0.30 &  0.10 &  0.16 &  0.19  \\
  BD+37 1458 & 5332 &  3.13 & $-$2.01 &  0.31 &  0.00 &  0.30 &  0.10 &  0.16 &  0.19  \\
      HD 6755 & 5075 &  2.41 & $-$1.63 &  0.03 & $-$0.25 & $-$0.00 &  0.10 &  0.16 &  0.19  \\
     HD 44007 & 4773 &  1.68 & $-$1.89 &  0.39 & $-$0.18 & $-$0.00 &  0.10 &  0.16 &  0.19  \\
     HD 63791 & 4619 &  1.34 & $-$1.87 &  0.12 & $-$0.29 & $-$0.10 &  0.10 &  0.16 &  0.19  \\
     HD 74462 & 4510 &  1.11 & $-$1.76 &  0.31 & $-$0.34 & $-$0.20 &  0.10 &  0.16 &  0.19  \\
     HD 84937 & 6312 &  4.51 & $-$1.91 &  0.21 & $-$0.07 & $<$ 0.10 &  \nodata &  0.14 &  \nodata  \\
     HD 94028 & 5925 &  4.63 & $-$1.36 &  0.36 & $-$0.04 & $-$0.05 &  0.10 &  0.14 &  0.17  \\
    HD 115444 & 4775 &  1.68 & $-$2.86 &  0.26 &  0.34 & $-$0.15 &  0.10 &  0.16 &  0.19  \\
    HD 122563 & 4610 &  1.32 & $-$2.54 & $-$0.06 & $-$1.01 & $-$0.40 &  0.10 &  0.16 &  0.19  \\
    HD 163810 & 5392 &  4.73 & $-$1.31 &  0.11 &  0.09 & $-$0.50 &  0.10 &  0.14 &  0.17  \\
    HD 204543 & 4570 &  1.24 & $-$1.98 &  0.53 &  0.23 & $-$0.45 &  0.10 &  0.16 &  0.19  \\
 BS 16076-006 & 5614 &  3.51 & $-$3.00 & $<-$1.50 & $<-$0.56 &  0.60 &  0.20 &  0.16 &  0.26  \\
 BS 16077-007 & 5958 &  3.68 & $-$2.82 &  0.27\tablenotemark{\dag} & $<-$0.58 &  1.00 &  0.20 &  0.16 &  0.26  \\
 BS 16080-054 & 4838 &  1.83 & $-$2.74 &  0.59 & $-$0.19 & $-$0.50 &  0.20 &  0.16 &  0.26  \\
 BS 16080-093 & 5120 &  2.53 & $-$2.73 &  0.08 &  0.11 & $-$0.50 &  0.20 &  0.16 &  0.26  \\
 BS 16084-160 & 4762 &  1.66 & $-$2.93 & $-$1.97 & $<-$1.71 & $-$0.15 &  0.20 &  0.16 &  0.26  \\
 BS 16085-050 & 4882 &  1.93 & $-$2.88 & $-$1.72 & $<-$1.35 & $-$0.80 &  0.20 &  0.16 &  0.26  \\
 BS 16467-062 & 5219 &  2.80 & $-$3.79 & $<-$1.21 & $<-$0.09 &  0.30 &  0.30 &  0.16 &  0.34  \\
 BS 16470-092 & 5948 &  3.67 & $-$1.83 &  0.23 & $-$0.42 &  0.20 &  0.20 &  0.16 &  0.26  \\
 BS 16472-018 & 4946 &  2.08 & $-$2.46 &  0.16 & $-$0.09 & $-$0.25 &  0.20 &  0.16 &  0.26  \\
 BS 16472-081 & 4835 &  1.82 & $-$1.56 &  0.16 & $-$0.06 & $-$0.15 &  0.20 &  0.16 &  0.26  \\
 BS 16477-003 & 4919 &  2.02 & $-$3.12 &  0.12 & $-$0.32 &  0.30 &  0.10 &  0.16 &  0.19  \\
 BS 16543-092 & 4523 &  1.14 & $-$2.36 & $-$0.39 & $-$0.58 & $-$1.00 &  0.20 &  0.16 &  0.26  \\
 BS 16546-076 & 4775 &  1.68 & $-$1.74 &  0.39 &  0.10 & $-$0.15 &  0.10 &  0.16 &  0.19  \\
 BS 16547-005 & 4917 &  2.01 & $-$1.79 &  0.04 & $-$0.15 & $-$0.10 &  0.10 &  0.16 &  0.19  \\
 BS 16547-006 & 6047 &  3.72 & $-$2.43 & $-$0.47 & $-$0.43 &  0.65 &  0.20 &  0.16 &  0.26  \\
 BS 16547-025 & 5197 &  2.74 & $-$1.62 &  0.37 &  0.48 & $-$0.80 &  0.20 &  0.16 &  0.26  \\
 BS 16547-099 & 5856 &  3.63 & $-$1.34 &  0.44 &  0.68 &  0.20 &  0.15 &  0.16 &  0.22  \\
 BS 16550-087 & 4791 &  1.72 & $-$3.10 &  0.80 & $-$0.98 & $-$0.65 &  0.20 &  0.16 &  0.26  \\
 BS 16551-058 & 4937 &  2.06 & $-$2.01 &  0.41 &  0.31 & $-$0.80 &  0.20 &  0.16 &  0.26  \\
 BS 16920-017 & 4854 &  1.87 & $-$3.02 & $-$0.48 & $<-$1.37 & $-$0.30 &  0.20 &  0.16 &  0.26  \\
 BS 16928-053 & 4743 &  1.61 & $-$2.87 &  0.12 & $-$0.75 & $-$0.15 &  0.15 &  0.16 &  0.22  \\
 BS 16929-005 & 5212 &  2.78 & $-$3.27 &  0.52 & $-$0.20 &  1.05 &  0.15 &  0.16 &  0.22  \\
 BS 16934-009 & 4256 &  0.59 & $-$1.55 &  0.30 &  0.22 & $-$0.60 &  0.20 &  0.16 &  0.26  \\
 BS 16934-072 & 6187 &  3.78 & $-$1.80 & $-$0.20 &  0.29 &  0.40 &  0.20 &  0.16 &  0.26  \\
 BS 16945-054 & 5281 &  2.99 & $-$2.93 & $-$0.32 & $-$0.07 &  0.10 &  0.20 &  0.16 &  0.26  \\
 BS 16972-003 & 6231 &  3.80 & $-$2.42 & $-$0.18 & $-$0.30 & $<$ 0.65 &  \nodata &  0.16 &  \nodata  \\
 BS 16972-013 & 5715 &  3.56 & $-$1.96 &  0.16 &  0.11 & $-$0.04 &  0.20 &  0.16 &  0.26  \\
 BS 16981-009 & 5259 &  2.92 & $-$2.88 & $-$0.77 &  0.06 &  0.40 &  0.15 &  0.16 &  0.22  \\
 BS 16986-072 & 4478 &  1.04 & $-$1.53 &  0.03 & $-$0.19 & $-$0.45 &  0.15 &  0.16 &  0.22  \\
 BS 17139-007 & 5918 &  3.66 & $-$2.42 & $-$0.58 & $-$0.57 &  0.30 &  0.20 &  0.16 &  0.26  \\
 BS 17444-032 & 5960 &  3.68 & $-$2.43 & $-$0.07 & $-$0.44 & $<$ 0.35 &  \nodata &  0.16 &  \nodata  \\
 BS 17446-025 & 5960 &  3.68 & $-$2.38 & $-$0.52 & $-$0.82 & $<$ 0.50 &  \nodata &  0.16 &  \nodata  \\
 BS 17570-090 & 5924 &  3.66 & $-$2.60 & $-$0.80 & $<-$1.72 & $<$ 0.30 &  \nodata &  0.16 &  \nodata  \\
 BS 17576-002 & 6203 &  3.78 & $-$1.38 &  0.38 & $-$0.09 &  0.40 &  0.10 &  0.16 &  0.19  \\
 BS 17576-071 & 6200 &  3.79 & $-$1.72 &  0.17 & $-$0.12 &  0.15 &  0.20 &  0.16 &  0.26  \\
 BS 17585-080 & 4630 &  1.36 & $-$1.38 & $-$0.12 & $-$0.53 & $-$1.00 &  0.15 &  0.16 &  0.22  \\
 CS 22166-016 & 5388 &  3.26 & $-$2.47 &  0.27 & $-$0.19 &  0.25 &  0.10 &  0.16 &  0.19  \\
 CS 22174-012 & 4934 &  2.06 & $-$2.52 & $-$0.48 & $-$0.90 & $-$0.70 &  0.20 &  0.16 &  0.26  \\
 CS 22180-005 & 5851 &  3.63 & $-$1.99 &  0.19\tablenotemark{\dag} &  0.03 &  0.30 &  0.10 &  0.16 &  0.19  \\
 CS 22183-015 & 5178 &  2.69 & $-$3.17 &  1.22\tablenotemark{\dag} &  1.77 &  2.10 &  0.20 &  0.16 &  0.26  \\
 CS 22183-031 & 5416 &  3.31 & $-$2.71 &  0.16 &  0.46 &  0.35 &  0.10 &  0.16 &  0.19  \\
 CS 22185-007 & 5193 &  2.73 & $-$2.45 & $-$0.45\tablenotemark{\dag} & $-$0.46 &  0.25 &  0.10 &  0.16 &  0.19  \\
 CS 22190-007 & 6013 &  3.71 & $-$2.02 &  0.62 &  0.91 &  0.00 &  0.20 &  0.16 &  0.26  \\
 CS 22878-101 & 4789 &  1.72 & $-$2.93 & $-$0.07 & $-$0.53 & $-$0.45 &  0.10 &  0.16 &  0.19  \\
 CS 22880-086 & 5457 &  3.37 & $-$2.42 &  0.08 & $-$0.74 &  0.20 &  0.20 &  0.16 &  0.26  \\
 CS 22883-020 & 6099 &  3.74 & $-$2.02 & $-$0.38 & $-$0.75 & $<$ 0.40 &  \nodata &  0.16 &  \nodata  \\
 CS 22892-052 & 4854 &  1.87 & $-$2.99 &  0.79 &  1.32 &  1.00 &  0.10 &  0.16 &  0.19  \\
 CS 22893-010 & 5528 &  3.44 & $-$2.38 & $-$0.02\tablenotemark{\dag} & $<-$1.00 &  0.30 &  0.20 &  0.16 &  0.26  \\
 CS 22898-027 & 5750 &  3.58 & $-$2.29 &  1.12 &  2.16 &  1.45 &  0.10 &  0.16 &  0.19  \\
 CS 22944-032 & 5528 &  3.44 & $-$2.64 &  0.44\tablenotemark{\dag} & $-$0.58 &  0.65 &  0.20 &  0.16 &  0.26  \\
 CS 22949-029 & 6244 &  3.81 & $-$1.70 & $-$0.45\tablenotemark{\dag} & \nodata &  0.55 &  0.20 &  0.16 &  0.26  \\
 CS 22949-048 & 4828 &  1.81 & $-$2.90 & $-$1.10 & $<-$1.38 & $-$0.20 &  0.10 &  0.16 &  0.19  \\
 CS 22950-153 & 5293 &  3.02 & $-$2.10 &  0.25 & $-$0.25 & $-$0.15 &  0.20 &  0.16 &  0.26  \\
 CS 22957-022 & 5075 &  2.41 & $-$3.02 & $-$0.28 & $-$0.89 &  0.30 &  0.15 &  0.16 &  0.22  \\
 CS 22962-006 & 6325 &  4.51 & $-$2.25 & $<-$1.42 & $<-$0.43 & $<$ 0.50 &  \nodata &  0.14 &  \nodata  \\
 CS 22963-004 & 5803 &  3.61 & $-$3.03 & $<-$0.97 & $<-$0.27 & $<$ 0.95 &  \nodata &  0.16 &  \nodata  \\
 CS 22965-016 & 4904 &  1.99 & $-$2.59 &  0.09\tablenotemark{\dag} & $<-$1.07 & $-$0.70 &  0.20 &  0.16 &  0.26  \\
 CS 22965-029 & 5467 &  3.38 & $-$2.31 &  0.61 &  0.16 & $-$0.60 &  0.20 &  0.16 &  0.26  \\
 CS 29497-040 & 5487 &  3.41 & $-$2.80 & $<-$1.35 & $<-$0.89 &  0.35 &  0.20 &  0.16 &  0.26  \\
 CS 29502-092 & 5114 &  2.51 & $-$2.92 & $-$0.18 & $-$0.92 &  1.15 &  0.10 &  0.16 &  0.19  \\
 CS 29510-008 & 5577 &  3.48 & $-$1.69 &  0.14 & $-$0.24 &  0.15 &  0.10 &  0.16 &  0.19  \\
 CS 29510-058 & 5192 &  2.73 & $-$2.51 &  0.26 & $-$0.14 &  0.50 &  0.10 &  0.16 &  0.19  \\
 CS 29517-025 & 5647 &  3.53 & $-$2.19 &  0.74\tablenotemark{\dag} &  0.71 & $-$0.35 &  0.20 &  0.16 &  0.26  \\
 CS 30306-082 & 5598 &  3.50 & $-$2.60 & $-$0.25\tablenotemark{\dag} & $-$0.41 &  0.25 &  0.20 &  0.16 &  0.26  \\
 CS 30306-110 & 6056 &  3.72 & $-$1.85 & $-$0.85\tablenotemark{\dag} & $-$0.65 &  0.05 &  0.25 &  0.16 &  0.30  \\
 CS 30306-126 & 5812 &  3.61 & $-$1.93 & $-$0.67 & $-$0.58 & $-$0.05 &  0.20 &  0.16 &  0.26  \\
 CS 30311-050 & 5710 &  3.56 & $-$1.51 & $-$0.09 & $-$0.12 &  0.15 &  0.15 &  0.16 &  0.22  \\
 CS 30315-093 & 5638 &  3.52 & $-$2.56 & $-$0.14 & $-$0.52 & $-$0.05 &  0.20 &  0.16 &  0.26  \\
 CS 30320-006 & 5661 &  3.53 & $-$1.89 & $-$0.11 & $-$0.61 & $-$0.10 &  0.20 &  0.16 &  0.26  \\
 CS 30320-069 & 6119 &  3.75 & $-$2.02 & $-$0.58\tablenotemark{\dag} & $<-$0.62 &  0.20 &  0.20 &  0.16 &  0.26  \\
 CS 30325-028\tablenotemark{*} & 4887 &  1.95 & $-$2.79 &  0.36 & $-$0.43 &  0.55 &  0.10 &  0.16 &  0.19  \\
 CS 30329-078 & 5475 &  3.39 & $-$1.77 &  0.12 &  0.09 &  0.05 &  0.10 &  0.16 &  0.19  \\
 CS 30329-129 & 5467 &  3.37 & $-$2.41 &  0.01 & $-$0.60 &  0.15 &  0.15 &  0.16 &  0.22  \\
 CS 30331-018 & 5379 &  3.24 & $-$1.77 &  0.32 &  0.02 &  0.00 &  0.20 &  0.16 &  0.26  \\
 CS 30332-114 & 5851 &  3.63 & $-$1.85 & $-$0.25 & $-$0.31 &  0.15 &  0.20 &  0.16 &  0.26  \\
 CS 30338-119 & 5611 &  3.50 & $-$1.82 & $-$0.33 & $-$0.60 &  0.05 &  0.20 &  0.16 &  0.26  \\
 CS 31060-030 & 5682 &  3.55 & $-$1.68 &  0.13 & $-$0.83 &  0.15 &  0.15 &  0.16 &  0.22  \\
 CS 31060-043 & 5452 &  3.36 & $-$2.06 &  0.36 &  0.79 &  0.05 &  0.15 &  0.16 &  0.22  \\
 CS 31061-062 & 5465 &  3.38 & $-$2.62 & $-$0.13 & $-$0.22 & $<-$0.25 &  \nodata &  0.16 &  \nodata  \\
 CS 31062-050 & 5313 &  3.08 & $-$2.65 &  1.40 &  2.37 &  1.70 &  0.20 &  0.16 &  0.26  \\
 CS 31069-064 & 5468 &  3.38 & $-$2.19 & $-$0.56 & $-$0.90 &  0.20 &  0.15 &  0.16 &  0.22  \\
 CS 31070-058 & 4864 &  1.89 & $-$2.29 &  0.14 & $-$0.21 &  0.15 &  0.10 &  0.16 &  0.19  \\
 CS 31078-018 & 5106 &  2.49 & $-$3.02 &  0.12 &  0.13 &  0.60 &  0.10 &  0.16 &  0.19  \\
 CS 31082-001 & 4893 &  1.96 & $-$2.78 &  0.68 &  1.22 &  0.30 &  0.10 &  0.16 &  0.19  \\
 CS 31085-024 & 4931 &  2.05 & $-$3.30 & $-$0.95 & $<-$1.32 &  0.05 &  0.15 &  0.16 &  0.22  \\
 CS 31088-083 & 5386 &  3.26 & $-$1.99 &  0.19 & $-$0.06 &  0.10 &  0.10 &  0.16 &  0.19  
\enddata
\tablenotetext{*}{The value of [Ba/Fe] was incorrectly given as $-$1.62 in Paper I.}
\tablenotetext{\dag}{The error from the synthetic fit on these objects
  was $\sim$0.35 dex}
\end{deluxetable}

\clearpage

\begin{deluxetable}{lccc}
\tablecolumns{4}
\tablecaption{[C/Fe] comparisons with previous
  studies. \label{comparison}}
\tablehead{
\colhead{Star} & \colhead{[C/Fe]} & \colhead{[C/Fe]} & \colhead{} \\
\colhead{ID} & \colhead{this study} & \colhead{literature} &
\colhead{Ref.}
}
\startdata
BD+37 1458    &   0.30 &  0.14 & 1  \\
HD 6755        &   0.00 & $-$0.07 & 1  \\
HD 74462       &  $-$0.20 & $-$0.50 & 1  \\
HD 94028       &  $-$0.05 & $-$0.07 & 1  \\
HD 115444      &  $-$0.15 & $-$0.41 & 2  \\
HD 122563      &  $-$0.45 & $-$0.41 & 3  \\
HD 204543      &  $-$0.70 & $-$0.58 & 1  \\
BS 16080-054   &  $-$0.50 & $-0.10$ & 4 \\
BS 16084-160   &  $-$0.15 &  0.10 & 4 \\
BS 16467-062   &   0.30 &  0.25 & 3  \\
BS 16477-003   &   0.30 &  0.34 & 3  \\
BS 16928-053   &  $-$0.15 & $-$0.23 & 2 \\
BS 16929-005   &  1.05  &  0.92 & 2 \\
CS 22183-031   &  0.35 & 0.42 & 2 \\
CS 22878-101   &  $-$0.45 & $-$0.29 & 3  \\
CS 22892-052   &   1.00 &  0.89 & 3  \\
CS 22898-027   &  1.45 &  1.90 & 5 \\
CS 22949-048   &  $-$0.20 & 0.16 & 5 \\
CS 30325-028   &  0.55  &  0.60 & 4\\
CS 31082-001   &   0.30 &  0.21 & 3  
\enddata
\tablerefs{(1) \citet{gratton00}; (2) \citet{honda04}; (3) \citet{cayrel04}; (4) \citet{aoki05}; (5) \citet{mcwilliam95}}
\end{deluxetable}

\clearpage

\begin{deluxetable}{lccc}
\tablecolumns{4}
\tablecaption{[Sr/Fe] comparisons with previous
  studies. \label{comparisonsr}}
\tablehead{
\colhead{Star} & \colhead{[Sr/Fe]} & \colhead{[Sr/Fe]} & \colhead{} \\
\colhead{ID} & \colhead{this study} & \colhead{literature} &
\colhead{Ref.}
}
\startdata
HD 115444    & 0.26 &  0.04  &  1 \\
HD 122563    & $-$0.06  &   $-$0.26   &    2 \\
BS 16080-054 &  0.59  &    0.25   &    3 \\
BS 16084-160 & $-$1.97  &   $-$2.34   &    3 \\
BS16085-050  & $-$1.72 &    $-$1.71  & 1 \\
BS16928-053  & 0.12  &   $-$0.23   & 1 \\
BS16929-005  & 0.52  &    0.28   & 1 \\
CS22183-031  & 0.16  &  0.10  & 1 \\
CS 22878-101 & $-$0.07  &   $-$0.50   &    4 \\
CS 22892-052 &  0.79  &    0.63   &    5 \\
CS 22898-027 &  1.12  &    0.97   &    4 \\
CS 22949-048 & $-$1.10  &   $-$1.47   &    4 \\
CS 30325-028 &  0.36  &    0.27   &    3 \\
CS 31082-001 &  0.68  &    0.65   &    6 
\enddata
\tablerefs{(1) \citet{honda04}; (2) \citet{honda06}; (3) \citet{aoki05}; (4) \citet{mcwilliam95}; (5) \citet{sneden03}; (6) \citet{hill02}}
\end{deluxetable}

\clearpage

\begin{deluxetable}{lcccccc}
\tablecolumns{7}
\tablecaption{ Predicted Fraction of [C/Fe] $\geq 1$ Stars for Different Studies\label{cfedist}}
\tablehead{
\colhead{[C/Fe] dist.} & \colhead{BPS} & \colhead{error} & \colhead{F06} & \colhead{error}  
& \colhead{This Study} & \colhead{error} 
}
\startdata
\multicolumn{7}{c}{[Fe/H] $< -2.0$} \\
Case 1 & 0.14 & 0.02 & 0.15 & 0.03 & 0.12 & 0.05 \\
Case 2 & 0.15 & 0.01 & 0.19 & 0.02 & 0.08 & 0.04 \\
Case 3 & 0.12 & 0.01 & 0.15 & 0.02 & 0.07 & 0.04 \\
Case 4 & 0.20 & 0.02 & 0.21 & 0.03 & 0.18 & 0.06 \\
\multicolumn{7}{c}{[Fe/H] $< -2.5$} \\
Case 1 & 0.18 & 0.03 & 0.19 & 0.04 & 0.18 & 0.05 \\
Case 2 & 0.24 & 0.02 & 0.29 & 0.03 & 0.11 & 0.04 \\
Case 3 & 0.19 & 0.02 & 0.21 & 0.03 & 0.10 & 0.04 \\
Case 4 & 0.24 & 0.03 & 0.26 & 0.04 & 0.23 & 0.06 
\enddata
\end{deluxetable}

\clearpage

\end{document}